\documentclass[useAMS,usenatbib,usegraphicx]{mn2e}

 \usepackage{times}

\usepackage[dvips]{graphicx}
\usepackage[section]{placeins}
\title[Gravitational waves and nonaxisymmetric oscillations]{Gravitational 
waves and nonaxisymmetric oscillation modes in mergers of compact object binaries}
\author[N. Stergioulas, A. Bauswein, K. Zagkouris and H.-T. Janka]{Nikolaos 
Stergioulas$^{1}$, Andreas Bauswein$^{2}$, Kimon Zagkouris$^{1}$ and 
Hans-Thomas Janka$^{2}$ \\
$^{1}$Department of Physics, Section of Astrophysics, Astronomy and 
Mechanics Aristotle,
University of Thessaloniki, Thessaloniki, 54124 Greece\\
$^{2}$Max-Planck-Institut f\"ur Astrophysik, Karl-Schwarzschild-Str. 1, 
D-85748 Garching, Germany}
\begin{document}

\date{}

\pagerange{\pageref{firstpage}--\pageref{lastpage}} \pubyear{2002}

\maketitle

\label{firstpage}

\begin{abstract}
We study the excitation of nonaxisymmetric modes in the post-merger
phase of binary compact object mergers and the associated gravitational
wave emission. Our analysis is based on general-relativistic simulations, 
in the spatial conformal flatness approximation, 
using smoothed-particle-hydrodynamics for the evolution of matter, and we
use a set of equal and unequal mass models, described by two
nonzero-temperature hadronic equations of state and by one strange star 
equation of state. Through Fourier transforms of the evolution of matter 
variables, we can identify a number of oscillation modes, as well as several  
nonlinear components (combination frequencies). We focus on the dominant 
$m=2$ mode, which forms
a triplet with two nonlinear components that are the result of coupling
to the quasiradial mode. A corresponding triplet of frequencies is identified in the 
gravitational wave spectrum, when the individual masses of the compact objects 
are in the most likely range of 1.2 to 1.35 $M_\odot$. We can thus associate, 
through direct analysis of the dynamics of the fluid, a specific frequency
peak in the gravitational wave spectrum with the nonlinear component 
resulting from the difference between the $m=2$ mode and the quasi-radial mode.  
Once such observations become available, both the $m=2$ and quasiradial mode
frequencies could be extracted, allowing for the application of 
gravitational-wave asteroseismology to the post-merger
remnant and leading to tight constraints on the equation of 
state of high-density matter. 
\end{abstract}


\section{Introduction}
%The orbits of neutron-star binaries are known to shrink due to the emission of 
%gravitational waves (GWs) (see e.g. \citealt{2010ApJ...722.1030W}), which leads 
%to the merging of the binary components on the timescale of some hundred million 
%years. 
Mergers of binary compact objects are prime sources for 
second- and third-generation interferometric gravitational wave (GW) detectors
\citep{Abbott:2007kv2,Acernese:2006bj,2010CQGra..27q3001A}. The expected GW
 signals from such events are estimated through general-relativistic 
hydrodynamical simulations (see \citealt{2009arXiv0912.3529D} for a review). 
The outcome of these simulations depends on the binary parameters and
on the equation of state (EOS) of high-density matter. The latter is rather uncertain 
\citep{Lattimer2007109,Steiner2010} and currently it is unclear whether, for a given 
binary mass, the merger
would lead to a hypermassive compact object or to prompt collapse to a black hole. 
If a hypermassive compact object forms, it will not be axisymmetric, but it will
show transient nonaxisymmetric deformations, such as a bar-like shape, spiral arms, 
a double core structure and quasi\-radial and nonaxisymmetric oscillations of the matter.
Strong non-axisymmetric features should be distinguishable in the GW 
 signal and could be used for characterizing the hypermassive compact
object 
\citep{Zhuge1994,2002PhRvD..65j3005O,Shibata2002,2005PhRvD..71h4021S,Shibata2006,2007PhRvL..99l1102O,Baiotti2008,Kiuchi2009,2010PhRvD..81b4012B}.

% In the case of the formation of a hypermassive object, the GW signal of 
% neutron-star mergers consists of an inspiral phase and the ringdown of 
% the postmerger remnant (see Figure~bla). The inspiral signal is generated 
% by the orbital motion of the neutron stars, and in fact the GW emission 
% can be described very accurately by analytical methods within the point-particle 
% approximation until finite-size and hydrodynamical effects become important 
% a few milliseconds before the merging (see e.g. \citealt{lrr-2006-4}). On 
% the other hand, the mechanisms generating the postmerger signal are not yet 
% uncovered 
% , albeit the signal shows a rich diversity in terms of distinctive peaks in 
% the (power) spectrum (see Figures bla). A deeper understanding of this 
% late-time emission is interesting, 
% in particular because it may yield some insight to the unknown equation of 
% state of high-density matter. For instance, it is known that the most 
% pronounced peak in the power spectrum of the GW signal in the range of 
% 2 to 4 kHz depends sensitively on the equation of state 
% .

%So ein statement wie: It has even been proposed that the peak is 
%cause by the orbital motion of the two dense cores?

Here, we analyze the formed hypermassive compact object as an isolated 
gravitating fluid, studying its oscillation modes. Fourier transforms
of the evolved variables reveal that the fluid is oscillating in a number
of modes that have discrete frequencies throughout the star. 
%The mode frequencies are integer multiples of an $m=1$ mode. 
Furthermore, we
also identify several nonlinear components, sums and differences of 
discrete oscillation modes. The oscillations identified in the fluid are
in direct correspondence with peaks in the GW spectrum, as
obtained through the quadrupole formula. We focus on the main quadrupole ($m=2$) 
oscillation mode of the fluid, which appears as a triplet, the side bands
being due to the nonlinear coupling to the fundamental quasiradial ($m=0$) mode.
The lowest-frequency side band, the difference between the $m=2$ and $m=0$
frequencies, coincides with a peak in the GW 
spectrum that characterizes the 
merger phase in all models, in which the mass of both compact objects
are in the range of 1.2 to 1.35$M_\odot$. We thus propose that in the event of 
detection, one could extract 
both the $m=0$ and $m=2$ mode frequencies of the merged object, which could lead to 
tight constraints on the EOS of high-density matter. 
In effect, such a detection
would enable an analysis with GW asteroseismology of the 
remnants of binary compact object mergers. 

The simulation code is based on general relativistic 
smoothed-particle-hydrodynamics (SPH) and on a spatially-conformally-flat
spacetime approximation, as described in detail in \cite{2010PhRvD..81b4012B} 
and references therein. We simulate the merger of the compact objects following the 
evolution from the late inspiral phase through the merging and the 
formation of the hypermassive remnant, until a quasi-stationary state 
is reached.

The paper is structured as follows: In Sec. \ref{sec:id} we discuss the 
initial data used in our similations. In Sec. \ref{sec:num} we outline
the numerical methods used in the simulations and in the mode extraction
procedure. Section \ref{sec:res} presents our results in detail, while in 
Sec. \ref{sec:comp} we compare them to previous results in the literature. 
We conclude with Sec. \ref{sec:dis}.

\section[]{Initial Data}
\label{sec:id}
% In this section I assume that the next are going to be mentioned ( The EoS that 
% we used, the total number of the models for every EoS and the corresponding masses
%  of these models, why we picked those data)

%This section gives an overview of the equations of state used in this paper, the choice of our models, their dynamics and the features of the corresponding gravitational-wave signals.

We consider two different models of hadronic EOSs,
referred to as Shen \citep{1998NuPhA.637..435S} and LS \citep{1991NuPhA.535..331L} 
and a strange quark matter EOS, referred to as MIT60. Since during merging
temperatures of more than several ten MeV can be reached, all three EOSs 
take into account non-zero temperature effects
(see \citealt{2010PhRvD..82h4043B} for a discussion of the importance of temperature
effects in the merger context). 

The Shen EOS was derived within a relativistic mean-field theory 
and assumes an incompressibilty modulus for nuclear matter of $K=280$~MeV 
\citep{1998NuPhA.637..435S}.
Solving the relativistic equations of hydrostatic equilibrium, nonrotating neutron
stars described by the Shen EOS have radii of about 15~km
in the mass range of 0.5 to 1.7~$M_{\odot}$. The maximum mass of
nonrotating stars is 2.2~$M_{\odot}$. For the LS EOS a
liquid-drop model with $K=180$~MeV was adopted \citep{1991NuPhA.535..331L}, 
which yields neutron stars more compact in comparison to the Shen EOS
. Typical radii are of the order of 12~km for masses of 0.5 to 1.6~$M_{\odot}$. 
The LS EOS supports nonrotating objects with masses slightly above 1.8~$M_{\odot}$.

The MIT60 EOS follows from the strange matter
hypothesis \citep{PhysRevD.4.1601,PhysRevD.30.272}, i.e. that 3-flavor
quark matter of up, down and strange quarks is more stable than ordinary 
nuclear matter. In such a case, compact stars 
would be strange stars rather than neutron stars
(see e.g. \citealt{1996csnp.book.....G,2007ASSL..326.....H}), which has not
been strictly ruled out theoretically
or observationally at this point. 
For the MIT60 EOS the MIT Bag model \citep{Farhi:1984qu} was employed, with a bag constant of 
60$\mathrm{MeV/fm^3}$ (see \citealt{2010PhRvD..81b4012B} for details).  Strange star models 
are generally more compact than neutron star models of the same mass. Furthermore, 
the mass-radius relation of strange stars does not show the typical inverse 
relation of neutron stars. The mass-radius relation for all three EOSs included in our 
study can be found in Fig. 2 of \cite{2010PhRvD..81b4012B}.

We note that the recent discovery of a 2$M_{\odot}$ pulsar \citep{2010Natur.467.1081D} 
practically rules out EOSs which yield maximum masses of nonrotating compact stars below 
this limit. The maximum masses of nonrotating stars for the LS and MIT60 EOSs employed 
here fall short of this requirement, but not dramatically. 
Though to date several finite-temperature EOSs 
(\citealt{2010PhRvC..81a5803T, 2010NuPhA.837..210H, 2010PhRvC..82a5806S, 2010PhRvC..82d5802S, 2011PhRvC..83c5802S, 2011arXiv1103.5174S}) have been published, 
at the time of writing only those introduced above have been available to us and 
successfully incorporated in the simulation code. Thus, 
leaving out the LS and MIT60 EOSs would leave us with a single EOS, which would not test
the sensitivity of our results to the choice of EOS. We thus include the LS and MIT60 EOSs 
in our study for the sole purpose of estimating the sensitivity of our conclusions to the 
EOS employed, keeping in mind their disadvantage. Nevertheless, the individual stars that
make up the binaries have masses significantly lower than the maximum mass limit, and the
hypermassive object formed as a result of the merger is supported against collapse 
by strong differential rotation. Because of this, the outcome of the simulations with
the LS and MIT60 EOSs is still useful for qualitative comparisons to the cases where 
the Shen EOS is used. 
 
The masses of compact stars in binaries cluster at about 1.35~$M_{\odot}$ 
\citep{1999ApJ...512..288T,2010arXiv1010.5429Z}, which is also predicted by population 
synthesis studies (e.g.~\citealt{2008ApJ...680L.129B}). Therefore, we focus in our 
analysis on systems with two neutron stars with a gravitational mass of 1.35~$M_{\odot}$. 
In order to investigate unequal-mass mergers, we also consider 
configurations with a 1.2~$M_{\odot}$ neutron star and a more massive companion of 
1.35~$M_{\odot}$. For the MIT60 EOS, we include a configuration with two low-mass stars of 
1.1~$M_{\odot}$ each, an unequal-mass binary with 1.2~$M_{\odot}$ and 1.35~$M_{\odot}$, 
and a configuration with two stars of 1.35~$M_{\odot}$ each. In the latter case, however, 
the hypermassive object formed after merging only survives for a few dynamical timescales
before collapsing to a black hole, and we do not discuss this model in our study. 
Table~\ref{tab:models} lists all models considered further, where, for example, 
Shen 12135 refers to the simulation with a 1.2~$M_{\odot}$ and a 1.35~$M_{\odot}$ 
star employing the Shen EOS.

\section{Numerical Methods}
\label{sec:num}

\subsection{Simulations}

The hydrodynamical simulations of the binary merger and the postmerger remnant 
are performed with a three-dimensional general-relativistic SPH  
code. The Einstein field equations are solved within the condition of spatial 
conformal flatness, which requires the additional implementation of a GW backreaction 
scheme. For details of the code see \cite{2002PhRvD..65j3005O,2007A&A...467..395O}. 
The calculations start from 
a quasi-equilibrium orbit about three revolutions before the actual merger. For the 
neutron star models about 500,000 SPH particles are used, while the flat density profile 
of strange stars allows for a lower SPH particle resolution of about 130,000 particles. 
Simulations with a somewhat higher number of particles do not significantly affect 
the main features of our results. 
The simulations are carried out until a stable remnant in approximate rotational 
equilibrium has formed and oscillates for tens of dynamical timescales, or until 
the delayed collapse to a black hole occurs.

The dynamics of the models used in this study have been extensively discussed in 
\cite{2010PhRvD..81b4012B}, and the reader is referred to this publication for 
details. Here we only give a brief outline: 
While orbiting around the common center of mass, the stars approach each other 
increasingly faster due to angular momentum and energy losses by the GW emission. 
Prior to merging, tidal forces start to deform the stars. The strength of this 
effect depends on the EOS and on the masses of the binary components. 
The deformation is pronounced 
for the Shen EOS, but relatively small for the strange star models. 
Finally, the stars merge into a rotating double-core structure, where the two 
dense cores appear to bounce against each other a few times to ultimately form a 
differentially rotating hypermassive object with a single core. In the case
of unequal masses, such as our models Shen 12135 and LS 12135, the less massive binary 
component is tidally stretched and wrapped around the more massive one. Moreover, 
an extended spiral arm develops, feeding a dilute halo around the differentially 
rotating central object. In the case of the unequal mass strange star model MIT60 12135, 
the outcome is more similar to the equal mass case, because of the reduced tidal
effects. 

The equal-mass neutron star mergers also form thick disks around the 
remnants, because matter is shed off from the whole surface of the hypermassive 
objects. In contrast, the remnants of merging strange stars terminate at
a sharp surface, as did the initial stars. Matter is stripped off only later 
in the evolution when angular momentum transfer leads to the formation of two 
thin spiral arms, which then form a fragmented thin disk around the central 
object.  

The hypermassive remnants of the merger are supported against gravitational collapse
by differential rotation and thermal pressure. However, ongoing angular momentum 
transfer and loss by mass shedding finally 
leads to gravitational collapse and black hole formation. The delayed collapse 
can take up to hundeds of milliseconds for low-mass models, but our simulations 
were terminated at about 10 to 20~ms after merging. 

The GW emission of our models is analyzed by means of the quadrupole formula. 
In the low-frequency regime the spectra are not reliable 
because we simulate only a small fraction of the inspiral phase and thus 
miss most of the low-frequency part of the signal.

\begin{figure} 
\includegraphics[width=8.3cm]{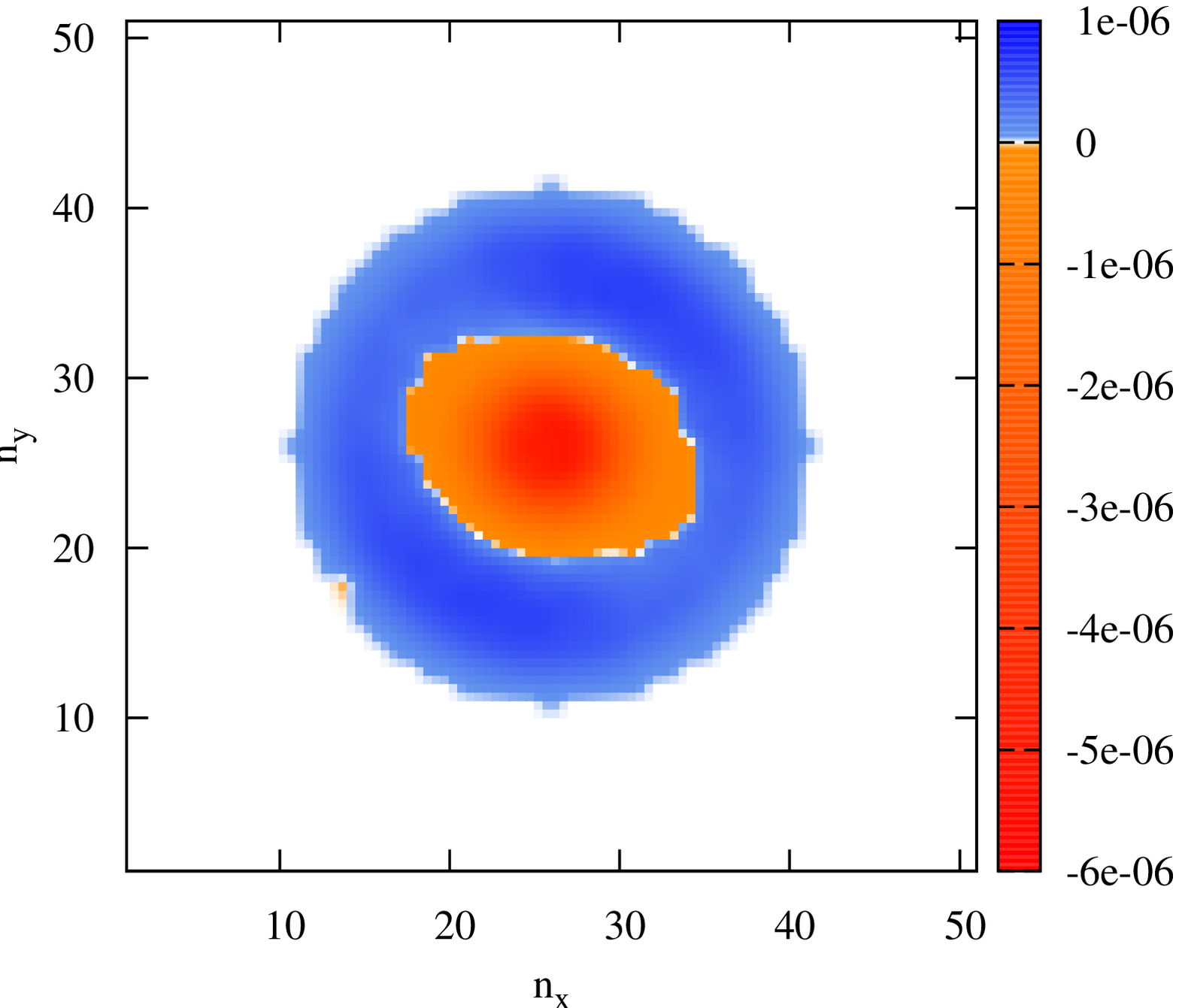}
\includegraphics[width=8.3cm]{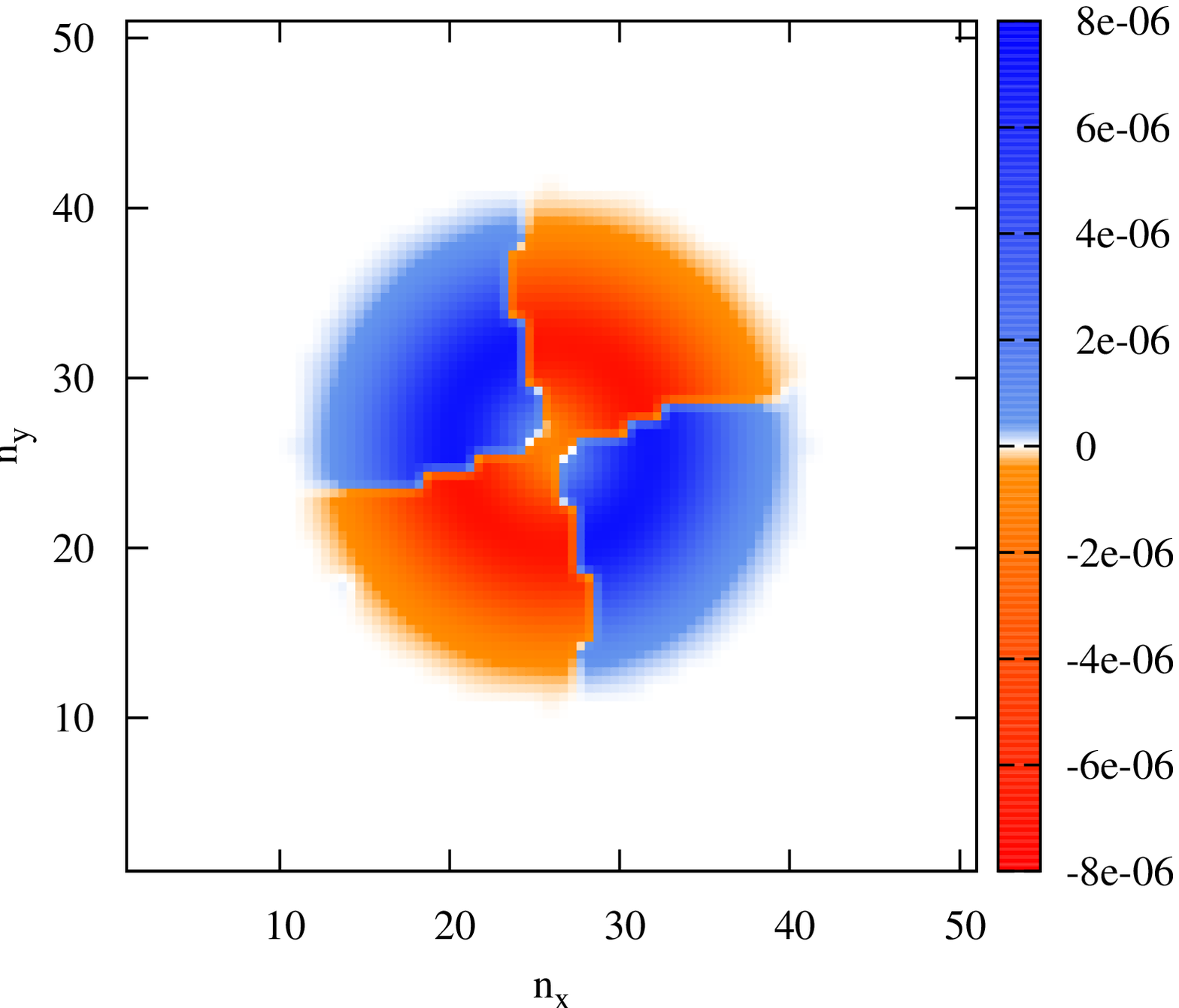}
\includegraphics[width=8.3cm]{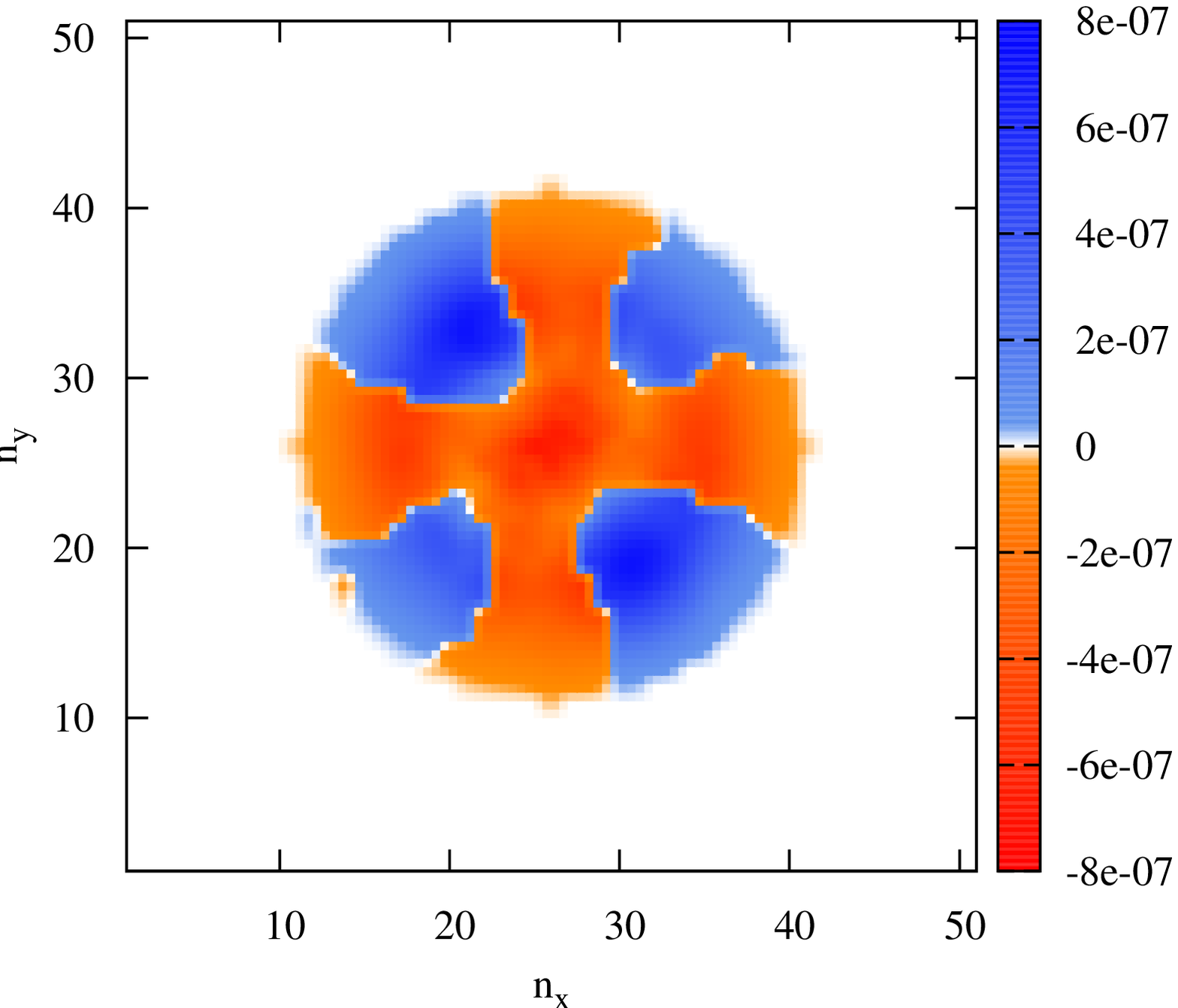}
  \caption{Projection in the equatorial plane of the extracted $m=0$, 2 and 4 mode eigenfuctions 
(from top to bottom) for the oscillations in pressure of model Shen 135135. The two axes 
count individual grid points of the Cartesian grid used for the mode analysis. The color scale 
only has a relative meaning.}
  \label{fig:2D-m0to4}
\end{figure}

% \begin{figure} 
% \includegraphics[width=8.3cm]{Figures-New/Shen-12135-2D-m1.eps}
%   \caption{ls12135 FFT of $p$ in equatorial plane.}
%   \label{fig:ls12135-p-fft}
% \end{figure}

\begin{figure} 
\includegraphics[width=8.3cm]{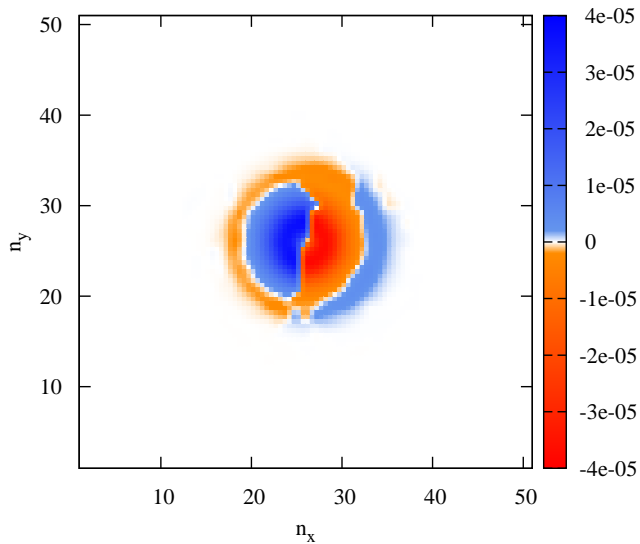}
  \caption{Projection in the equatorial plane of the extracted $m=1$ mode eigenfunction, 
 for the oscillations in pressure of model MIT60 12135.}
  \label{fig:2D-m1}
\end{figure}

% Sprachliche UNterscheidung NS <-> SS !!!??

% und irgendwo noch sagen, we have chosen exclusively configurations that form a hypermassive remnant instead of promptly collapsing to a black hole.
% 
% 
% 
% 2-solar-mass star erwaehnen?
% 
% 
% use "identification" of modes
% 
% mention toymodel in discussion/conclusions or motivation
% 
% Zhang paper mittlerweile akzeptiert

\subsection{Mode Extraction}

The original simulation data in different two-dimensional planes
 are first mapped onto a Cartesian grid of size 
$n_x \times n_y = 51\times 51$. We then perform Fourier
analysis of evolved variables over the whole grid (but focusing mainly on the 
equatorial plane) and identify the predominantly excited modes, which appear as 
discrete frequencies in the Fourier spectrum. At selected mode frequencies, we
proceed with extracting the shape of the mode eigenfunctions, using a method
described in \cite{Stergioulas04}. The two-dimensional shape of the eigenfunction 
in the equatorial plane correlates directly with the amplitude of the Fourier 
transform at a given mode frequency. At nodal lines, care must be taken to switch
the sign of the eigenfunction. This method was first applied in \cite{Stergioulas04}
to the case of axisymmetric modes, where two-dimensional eigenfunctions were
extracted in a plane that passes through the rotational axis. Here, we find that
the same method can be used for extracting the eigenfunctions of non-axisymmetric
modes in the equatorial plane. We note that for all modes that were identified, 
the mode frequency was discrete throughout the star and 
identical in all hydrodynamical variables. 

To our knowledge, this is the first 
direct demonstration that once a hypermassive object forms after a binary merger, 
it behaves as an isolated, self-gravitating system and its dynamics can be described
as a superposition of different oscillation modes (in a differentially rotating
background). Departures from a background equilibrium state can be 
described (up to a certain degree) by nonlinear features of the oscillations. 
We note that the background is varying in time, so that individual frequency
peaks are broadened. Nevertheless, the frequency peaks remain sufficiently narrow
to identify individual modes.

\section{Results}
\label{sec:res}

\subsection{Two-dimensional eigenfunctions}

Non-isentropic, differentially rotating, compact objects can 
oscillate in a large number of different quasinormal modes, ($f$-modes, $p$-modes, $g$-modes and
inertial modes, distinguished by three different mode numbers, $l$, $m$ and $n$, the latter
one being the radial order, see \cite{Stergioulas2003} for a review)). Here, 
we only characterize the structure of the 
extracted modes in the equatorial plane, in terms of the mode number $m$. 
A complete characterization and identification with specific quasinormal modes 
would require a much more detailed analysis that is out of the scope of the current work.
In particular, neither a linear, nor a nonlinear complete analysis of possible oscillation
modes has been done for remnants of mergers of compact objects, to date, with which we
could directly compare our results. 
As such, we will talk about ``modes'' and ``eigenfunctions'' even though we have not
strictly shown a correspondence to specific linear quasinormal modes. Nevertheless, for
the purpose of explaining the main observable features of GWs from nonaxisymmetric 
oscillations in binary compact object mergers, our present approach suffices. 

In the Fourier transform of hydrodynamical variables, the dominant mode is always
the $m=2$ mode. For equal-mass mergers, the even mode numbers are predominantly 
excited, while for unequal-mass mergers, the odd mode numbers are also considerably
excited (mainly the $m=1$ mode). The modes can be identified by inspection
of the extracted two-dimensional eigenfunction in the equatorial plane. 
Representative cases are shown in Figs. \ref{fig:2D-m0to4} and \ref{fig:2D-m1}, which 
display the eigenfunctions
(in pressure oscillations) of the $m=0$, 2 and 4 modes for the equal-mass model 
Shen 135135 and the $m=1$ mode for the unequal-mass model MIT60 12135. The $m=0$ 
(quasiradial) eigenfunction is dominated by a spherically symmetric contribution, while an 
additional, small (rotationally-induced) quadrupole distortion is also present. As
expected for this type of eigenfunction, there is a nearly circular nodal line 
in the pressure perturbations, characteristic of the fundamental radial mode.

The eigenfunction of the $m=2$ mode shows a very sharp quadrupole pattern, induced
by the merger of the two individual stars. The two nodal lines are, as expected, 
nearly perpendicular to each other. The $m=4$ mode eigenfunction shows a 
characteristic octupole pattern. The alternating regions are of nearly the same
size, but due to the very low amplitude of excitation and the finite duration of
the time-series, the Fourier transform also picks up a non-zero contribution at
the center of the star, which means that for this mode we are approaching the
accuracy of our method. Higher-order modes should also be excited (and there
are indeed a large number of peaks in the Fourier transform at high frequencies)
but their amplitude is so small that any extracted eigenfunction will be dominated
by numerical errors.

Fig. \ref{fig:2D-m1} shows the eigenfunction of the $m=1$ mode excited during the merger of
the uneqal-mass model MIT60 12135. A nodal line that cuts through the center of
the star is characteristic of this mode. Notice that 
conservation of momentum prohibits the existence of a fundamental $m=1$ mode, as this
would induce a linear motion of the center of mass. However, higher-order overtones
are allowed, so the extracted mode should be identified as the (lowest-frequency) 
first overtone of the $m=1$ modes. This is supported by the additional, nearly
circular node line, which allows for momentum conservation during the oscillation.

\subsection{Mode frequencies}

Table \ref{table:nonlin} lists the extracted mode frequencies (in the inertial frame, 
defined by coordinate time $t$) for the various models. 
We note that the frequencies of the quasiradial ($m=0)$ mode differ significantly 
among the three chosen EOSs. In the selected mass-range, this frequency is lowest 
for the Shen EOS (about 0.5kHz). It doubles to about 1kHz for the LS EOS and it 
reaches up to about 1.5kHz for the MIT60 EOS. This mode frequency is affected
mainly by two factors: the compactness of the star and the distance of the model
from the region of axisymmetric instability to collapse. At the boundary of this
region, the frequency of the quasi\-radial mode goes through zero. On the other hand, 
higher compactness leads to higher mode-frequency. But, because of the axisymmetric
instability, the quasi\-radial mode frequency can be small for any EOS, as long as
the model is near the instability threshold. This latter property explains the
fact that, for each EOS, the models with larger total mass have smaller quasi\-radial
mode frequency. 

The frequency of the $m=2$ mode depends mainly on the compactness of the star and 
for the models in Table \ref{table:nonlin} it is roughly 2 kHz for the stiffer 
Shen EOS and roughly 3kHz for the softer LS EOS and for the MIT60 EOS.  
As expected, for each EOS, it is higher for the more massive (hence more compact) model. 
In all unequal mass cases, we were able to extract the $m=1$ mode frequency, which
is induced by the unequal distribution of mass just prior to merging. The additional 
$m=3$ mode was extracted for the Shen and LS EOSs, while in three models we could
also identify the $m=4$ mode. 

A remarkable property of the nonaxisymmetric modes $m=1...4$ is that the extracted
frequencies are nearly integer multiples of the frequency of the $m=1$ mode. This 
property holds true with good accuracy for the equal-mass cases, while for 
the unequal-mass mergers it is less accurate, but still roughly true.

%  The high-frequency regime is dominated 
% by a pronounced peak. The positions $f_{peak}$ of this peak for the different 
% models are given in Table~bla. They confirm earlier 
% indings~\citep{2005PhRvD..71h4021S,2007PhRvL..99l1102O} that the value of 
% the ringdown frequency increases with the total mass of the binary. 
% Furthermore, a dependence on the equation of state has been reported 
% in~\citealt{2005PhRvD..71h4021S} and \citep{2007PhRvL..99l1102O}. As a 
% general observation, equations of state yielding more compact stars cause 
% higher peak frequencies, which is intuitively understandable, because 
% also the remnants are more compact and hence oscillate at higher frequencies. 
% Also this picture is supported by the values listed in Table~bla.
% 
% However, though its observational importance, the nature of these oscillations 
% has not yet been clarified. In order to link these frequencies to the physical 
% properties of the merger remnants and their equation of state, it is 
% interesting to reveal the oscillation mode that is responsible for the 
% signal peak.
% 
% 
% For this study we used a set of four EoS, namely Shen, Ls, MIT60 and MIT80. The 
% mode extraction procedure has been the same on all four EoS. In each case we study
%  the two main frequencies that correspond to the m=2 mode and to its first harmonic.
%  For every model we study the 2D eigenfunction profiles at these two specific frequencies
%  on both the polar and the equatorial plane. The data resulting from this analysis 
% are presented in the next subsections.

 \begin{table}
\caption{Extracted mode frequencies.}\label{tab:models}
\centering
\begin{tabular}{l ccccc}
\hline\hline
Model & $f_{m=0}$  & $f_{m=1}$ & $f_{m=2}$ & $f_{m=3}$ & $f_{m=4}$\\
& (kHz) & (kHz) & (kHz) & (kHz) & (kHz)\\
\hline
% Shen 12135 & $5.29\times 10^{-3}$ & $1.06\times 10^{-2}$ & 2.04\\
% Shen 135135 & $5.77\times 10^{-3}$ & $1.09\times 10^{-2}$ & 2.22\\
% Shen 1414 & $5.94\times 10^{-3}$ & $1.14\times 10^{-2}$ & 2.29\\
% LS 12135 & $7.62\times 10^{-3}$ & $1.53\times 10^{-2}$ & 3.04\\
% LS 135135 & $4.83\times 10^{-3}$ & $1.08\times 10^{-2}$ & 3.22\\
% MIT60 1111 & $7.19\times 10^{-3}$ & $1.47\times 10^{-2}$ & 2.85\\
% MIT60 1112 & $7.66\times 10^{-3}$ & $1.45\times 10^{-2}$ & 2.93\\
% MIT80 1111 & --- & $1.84\times 10^{-2}$ & 3.53\\
% MIT80 1112 & $1.01\times 10^{-3}$ & $1.95\times 10^{-2}$ & 3.86\\
Shen 12135 & 0.50 & 1.07 & 2.14 & 3.22 & 4.26 \\
Shen 135135 & 0.46 & -- & 2.24 & -- & 4.11 \\
LS 12135 & 1.10 & 1.55 & 3.12 & 4.66 & --\\
LS 135135 & 0.98 & -- & 3.30 & -- & --\\
MIT60 1111 & 1.56  & -- & 2.92 & --  &5.94 \\
MIT60 12135 & 1.24 & 1.72 & 3.26 & -- & -- \\
\hline
\end{tabular}
\label{table:nonlin}

\end{table}

\subsection{A representative model}

\begin{figure}
\includegraphics[width=8.1cm]{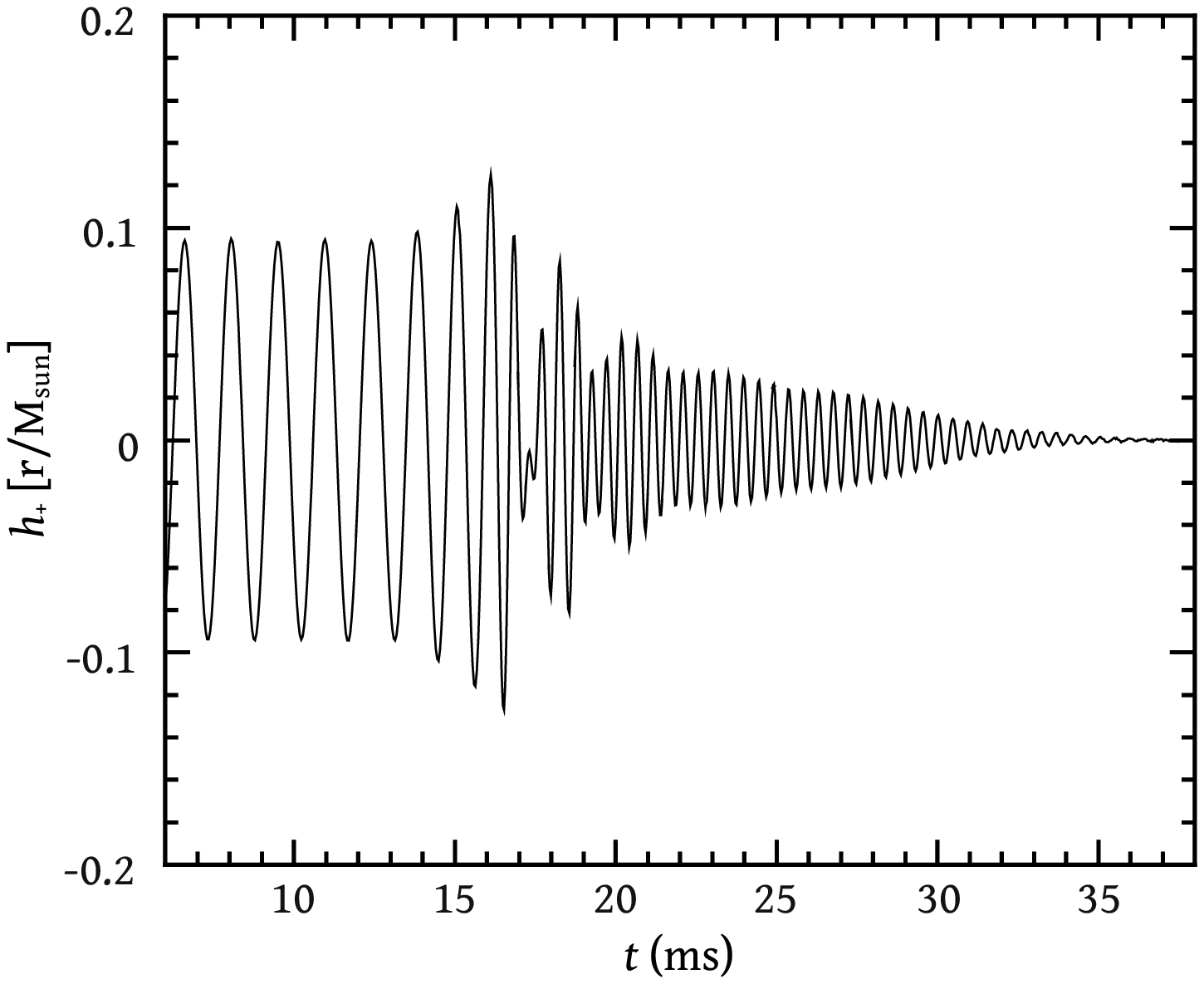} 
\includegraphics[width=8.1cm]{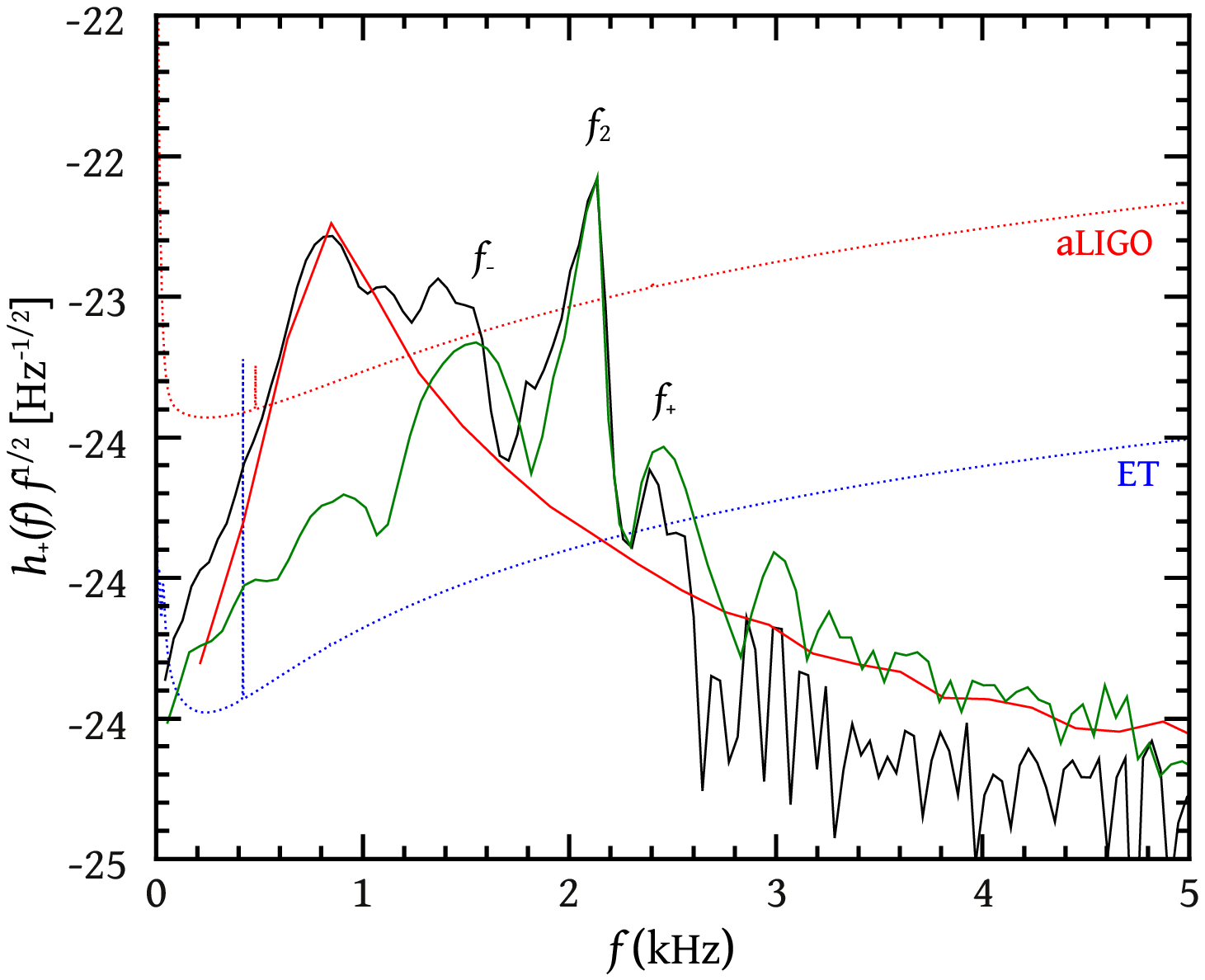}
\includegraphics[width=8.1cm]{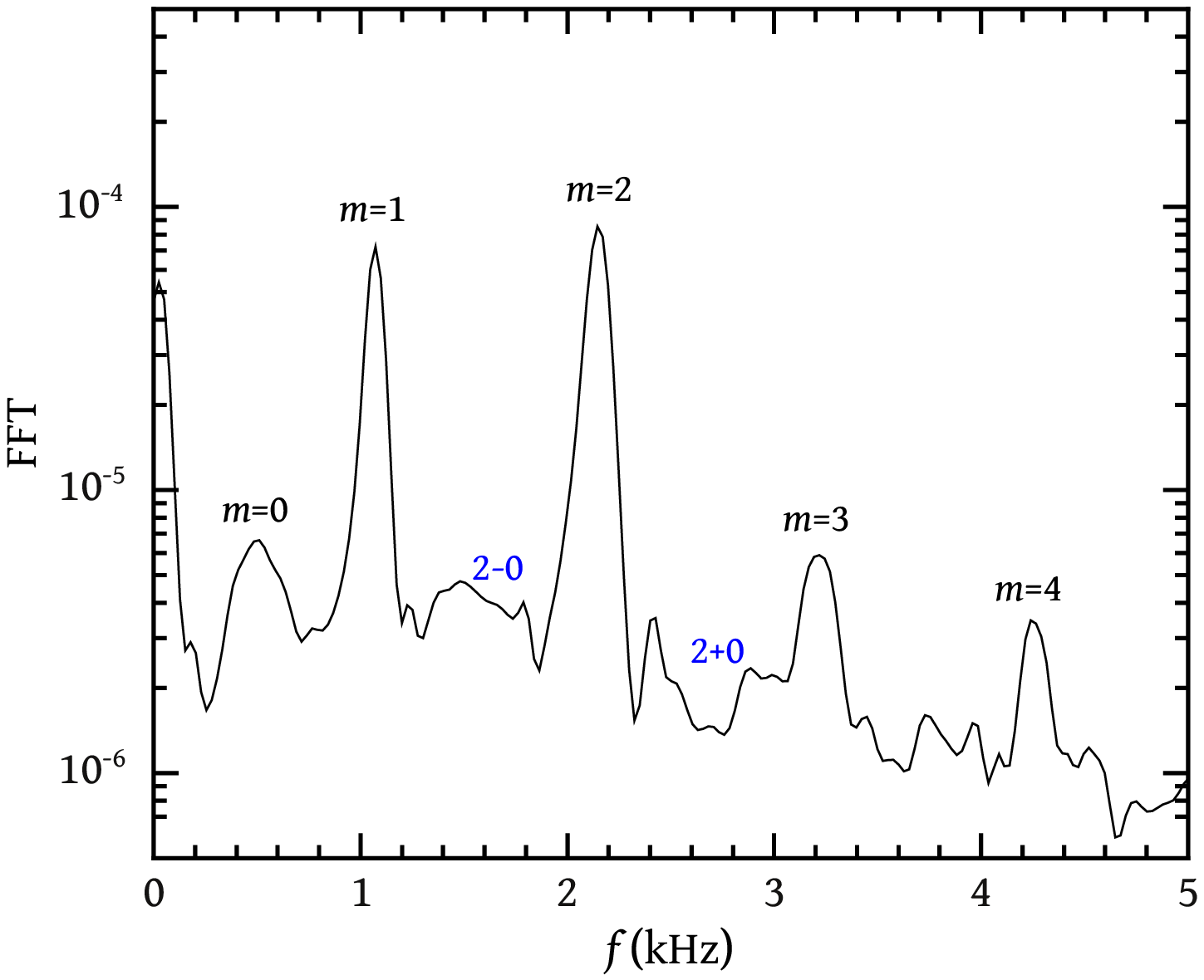}
  \caption{Model Shen 12135 -- Top panel: Time evolution of the GW amplitude $h_{+}$. 
Middle panel: Total (black), pre-merger (red) and post-merger (green) 
scaled power spectral density, compared to the Advanced LIGO and ET unity SNR sensitivity 
curves (\citet{2010CQGra..27h4006H,2010CQGra..27a5003H}). The distance to the source is assumed
to be 100 Mpc. 
Bottom panel: Amplitude of FFT for the time evolution of the pressure, $p$, in the
equatorial plane. Several oscillation modes, as well as nonlinear combination frequencies (blue labels)
are identified.}
  \label{fig:shen12135}
\end{figure}

The evolution of the GW amplitude $h_+$ expected from the late binary
inspiral, merger and post-merger phases for the unequal-mass model Shen 12135 is 
shown in the top panel of Fig. \ref{fig:shen12135}. The merger takes place at around 
$t=17$ms from the start of the simulation. A characteristic sudden change in the frequency and 
phase of the waveform takes place at the onset of merging. After a few highly nonlinear 
oscillations, the waveform settles into a more regular pattern. The late-time damping of the 
waveform is the result of a combination of shock-induced dissipation of oscillations 
and numerical damping. Gravitational-wave damping is effective only on larger timescales 
than shown here. With less numerical damping, the gravitational wave amplitude could
remain strong for a much longer time, leading to an enhanced signal (for a recent long-term 
evolution see \cite{Rezzolla2010}).

Irrespective of the true damping timescale, we are interested mainly 
in the frequency spectrum, which is shown in the middle panel of Fig. \ref{fig:shen12135}.
 We plot the scaled power spectral density, $h_+(f) \sqrt{f}$ (black curve), where $h_+(f)$ is the 
Fourier transform of $h_+(t)$, which is directly comparable with the anticipated sensitivity for
Advanced LIGO presently being installed (\cite{2010CQGra..27h4006H}) and the projected Einstein 
Telescope (ET) detectors (\cite{Punturo2010,2010CQGra..27a5003H}) (dotted red and blue curves, respectively). 
Note that the sensitivity of the advanced Virgo detector (\cite{Acernese:2006bj}) is planned to 
be similar to the one of the Advanced LIGO detector. Therefore, we include the latter only as an 
example for the class of detectors going into operation within the next years. The distance to 
the source is assumed to be 100 Mpc.
We also show the same quantity during the pre-merger phase only 
(red curve) and during the post-merger phase only (green curve). The first peak at roughly
900Hz is artificial, because our simulations only start at a few orbits before merging, so
most of the actual inspiral phase is missing. The largest peak (denoted by $f_2$) at about
2kHz is clearly produced exclusively in the post-merger remnant. In addition to this peak, 
which is very pronounced and clearly detectable by ET and with good prospects for detection
by Advanced LIGO, there are two more characteristic peaks, denoted by $f_-$ and $f_+$, which
are also produced in the post-merger phase and have good prospects for detection by ET 
(more so for $f_-$ than for $f_+$). These two peaks are both above the ET noise curve and 
above the contribution of the pre-merger (inspiral) waveform to the FFT of the signal. 
There are additional peaks above the ET noise curve in the range of 1kHz to 1.5kHz, which, 
however, will be superseded by the inspiral signal, unless the post-merger signal is analyzed 
separately from the inspiral part. Furthermore, the high-frequency peaks that are below 
the ET noise curve will not concern us. 

The three frequencies $f_-$, $f_2$ and $f_+$ form a triplet, which we 
will attempt to interpret in terms of the oscillation modes of the post-merger remnant.
The bottom panel of Fig. \ref{fig:shen12135} shows the amplitude of the Fourier transform of the 
evolution of the pressure along a fixed coordinate line in the equatorial plane which 
passes through the center of the coordinate system (this is an integrated amplitude, 
taking into account the contributions along the whole coordinate line). The Fourier
amplitude is composed of several very distinct and narrow peaks, which correspond to
the discrete $m=0,$ 1, 2, 3 and 4 mode frequencies. In addition, a number of smaller 
peaks are seen. The latter are nonlinear components, combination frequencies 
of the main oscillation modes,
i.e. linear sums and differences\footnote{These combination frequencies are also 
called bilinear coupling (or intermodulation) components.} (see \cite{Zanotti2005,Passamonti2007} 
and references therein). 
The difference of the frequencies of the
$m=2$ and $m=0$ mode is denoted as ``2-0'', their sum as ``2+0'' (and similarly
for other components). A two-dimensional plot of the Fourier amplitude of the pressure
evolution at the ``2-0'' frequency reveals that indeed it is a combination of 
a radial and a quadrupolar pattern. 

By comparison of the middle and bottom panels of Fig. \ref{fig:shen12135}, there
is an obvious near coincidence of the frequency peak $f_2$ in GWs 
with the frequency of the $m=2$ mode of the post-merger remnant. Furthermore, the
other two peaks of the GW triplet, $f_-$ and $f_+$, nearly coincide with
the ``2-0'' and ``2+0'' combinations of the pressure oscillations. As we
will see, this is a generic behaviour for all models (except for the very low mass 
MIT60 1111 model). It is thus tempting to attribute the $f_-$ peak in the 
GW spectrum to a nonlinear interaction between the quadrupole and 
quasiradial modes. Essentially, a double core structure that appears and disappears 
several times in the early post-merger phase could be the result of (or could
be described by) this nonlinear interaction. To firmly establish this goes beyond the 
scope of the present work, but already the coincidence of the $f_-$ and ``2-0'' frequencies 
has practical consequences: it allows the determination of both the $m=2$ and $m=0$ oscillation
frequencies of the post-merger remnant. In combination with the additional knowledge
of the characteristics of the compact objects from the detection of the inspiral
signal (see \cite{Read2009} and references therein), the determination of at least two 
independent oscillation frequencies should
allow for the application of the theory of GW asteroseismology, 
placing very stringent constraints on several properties of the post-merger
remnant and consequently on the EOS of high-density matter (for an application
of GW asteroseismology to isolated, rotating neutron stars, see 
\cite{Gaertig2011}).

\subsection{A survey of different models}

\begin{figure} 
  \includegraphics[width=8.1cm]{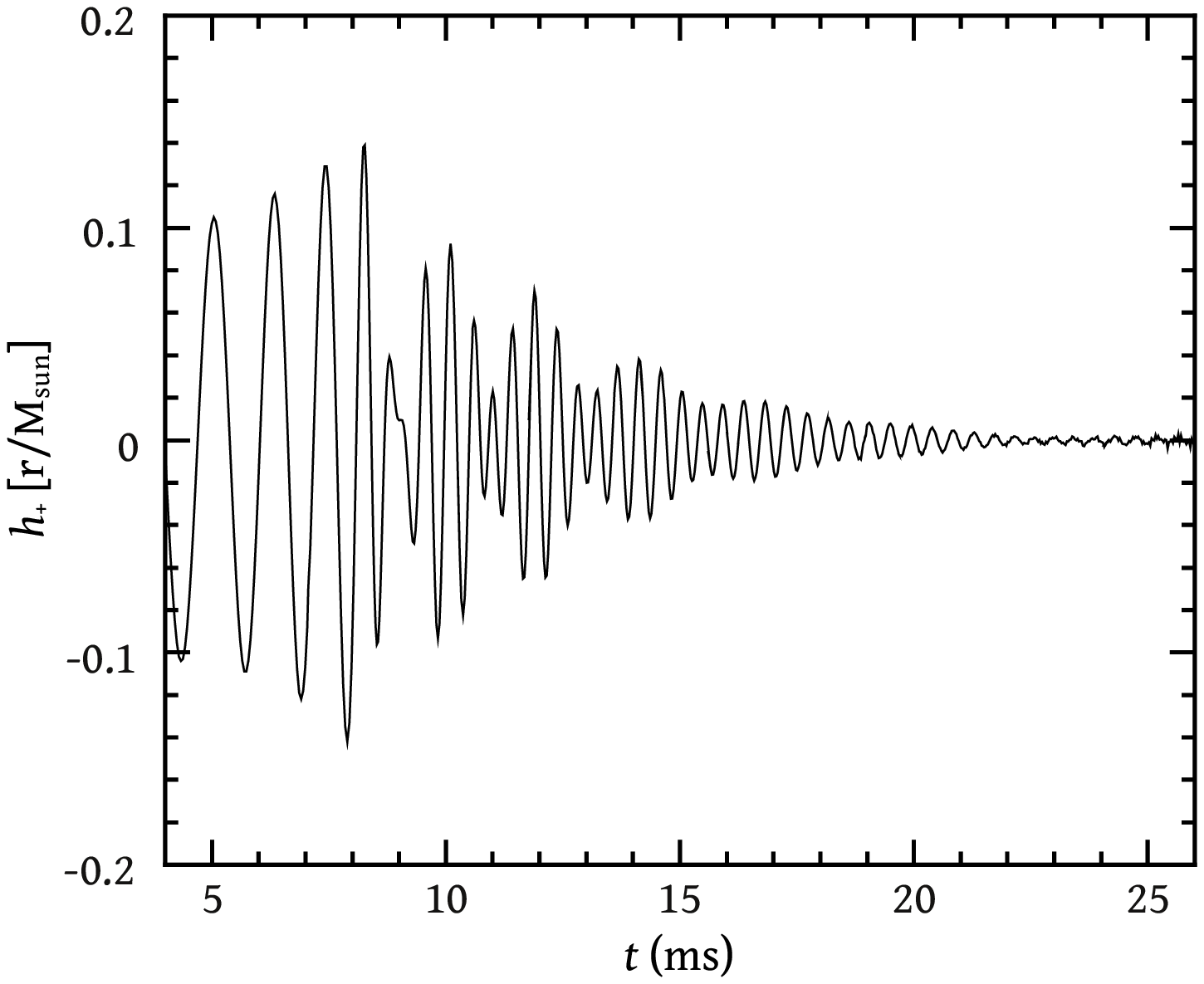}
 \includegraphics[width=8.1cm]{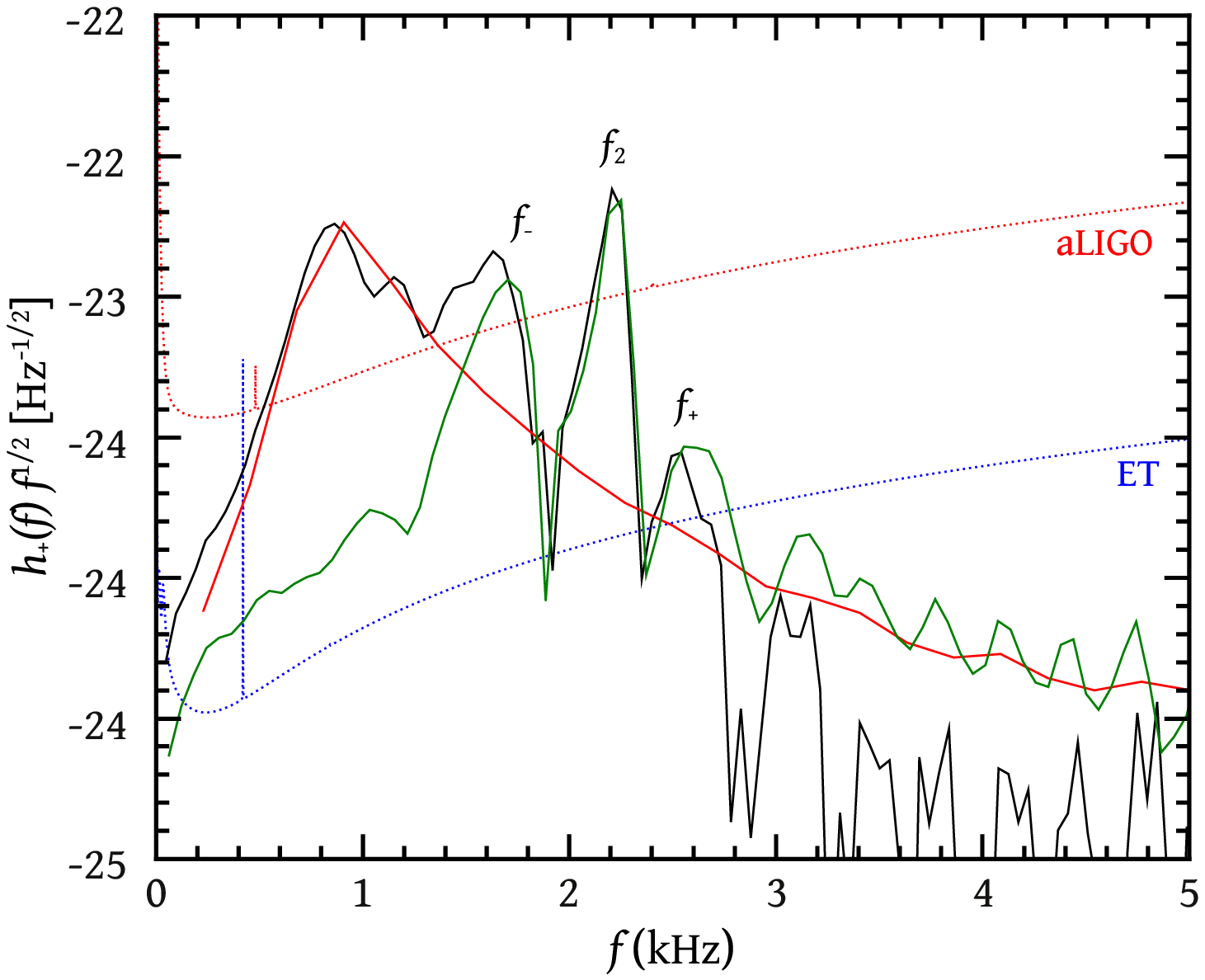}
 \includegraphics[width=8.1cm]{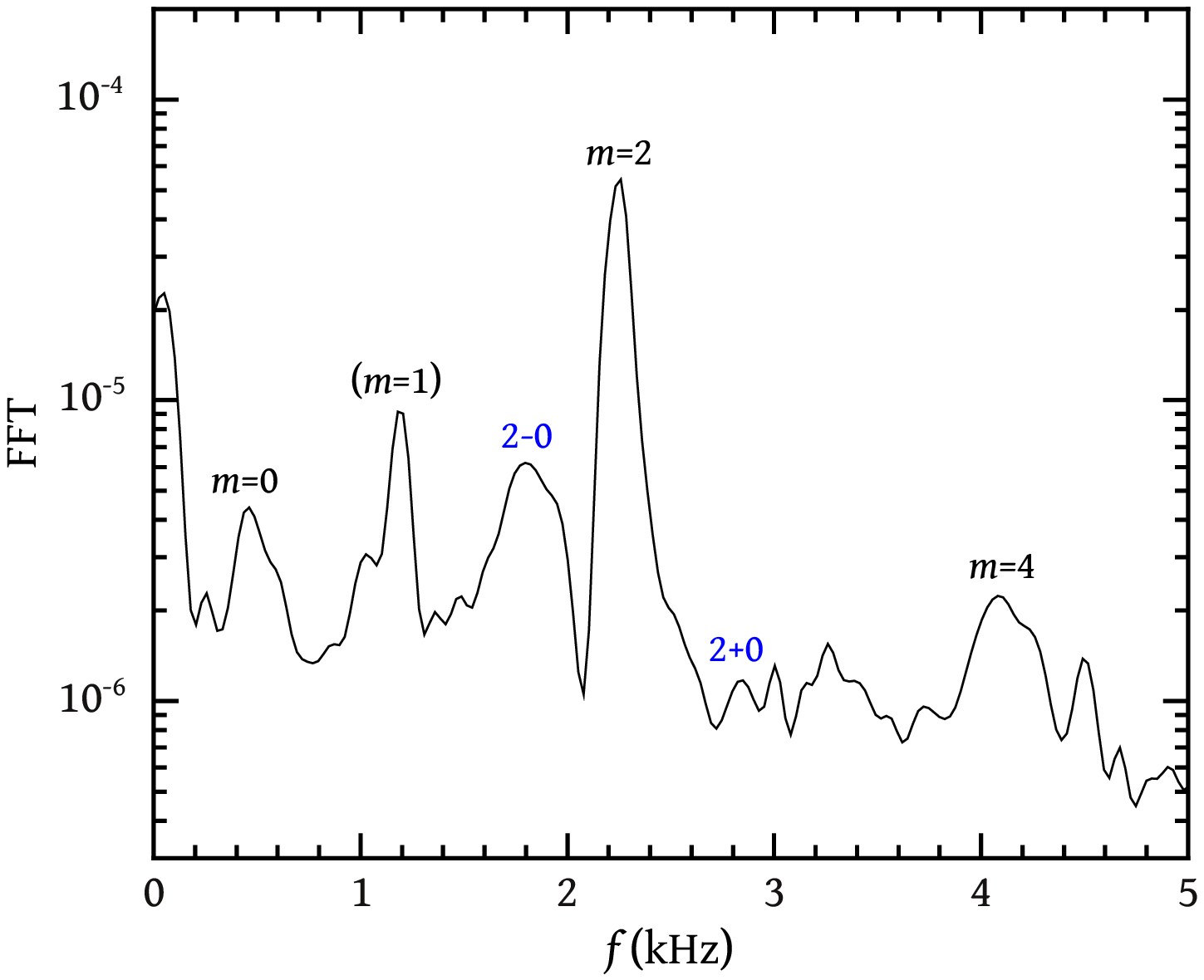}
  \caption{Same as Fig. \ref{fig:shen12135}, but for model Shen 135135.}
  \label{fig:shen135135}
\end{figure}

Having examined in detail model Shen 12135 as a representative case, it is instructive
to compare the main results to the equal-mass case Shen 135135 (Fig. \ref{fig:shen135135}).
The GW spectrum (middle panel of Fig. \ref{fig:shen135135}) is very similar
to the representative case, except for small differences in the frequencies of the main peaks.
In the Fourier transform of the pressure evolution (bottom panel of Fig. \ref{fig:shen135135}),
the odd modes $m=1$ and 3 are no longer as strong as in the unequal-mass case, and the evolution
is mainly determined by the $m=2$ mode, while the $m=0$ and $m=4$ modes are also present and so
are nonlinear components. Again, there is a coincidence between the frequency of the $f_-$ peak in the
GW spectrum and the ``2-0'' combination frequency in the pressure evolution. 

\begin{figure} 
  \includegraphics[width=8.1cm]{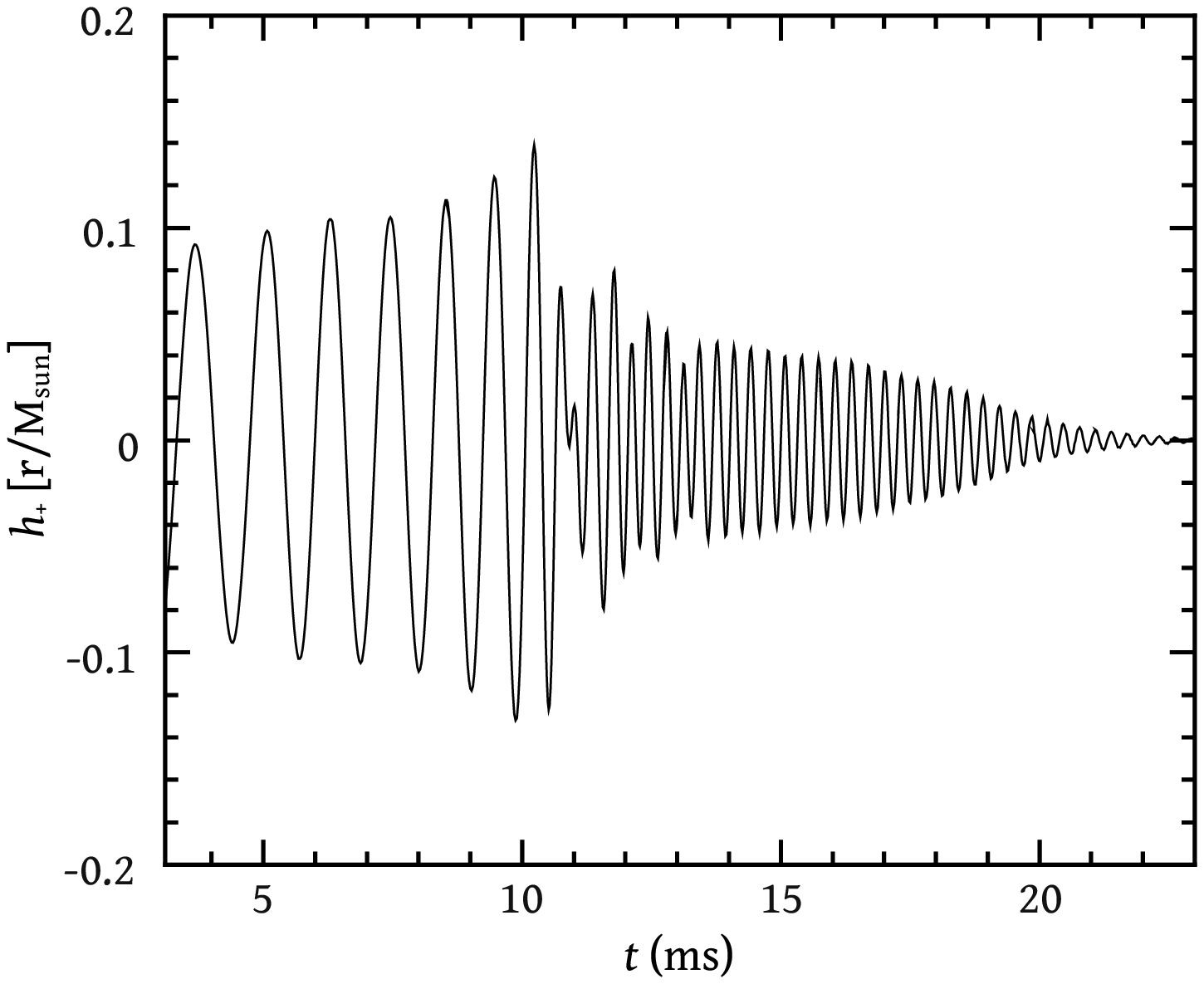}
 \includegraphics[width=8.1cm]{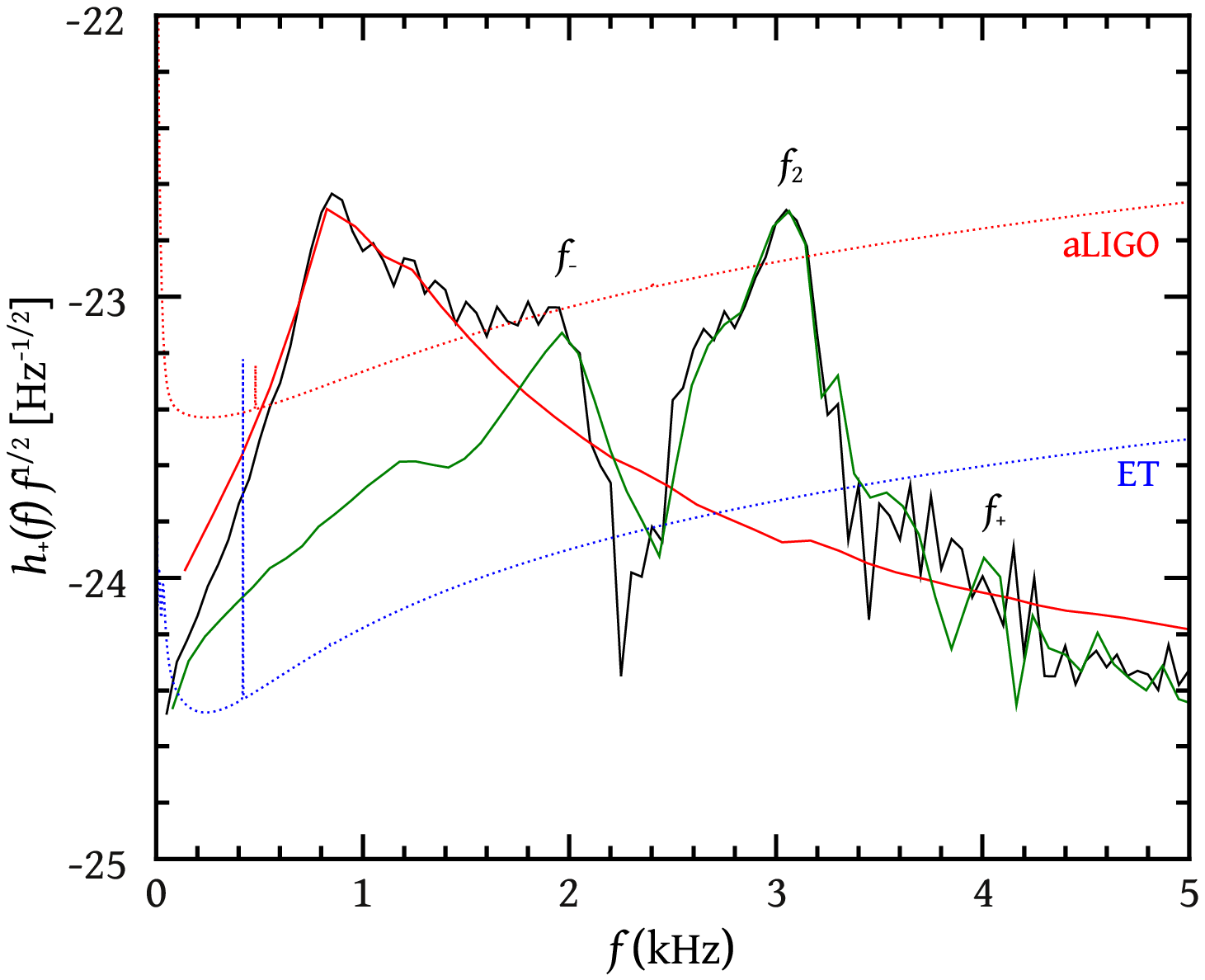}
\includegraphics[width=8.1cm]{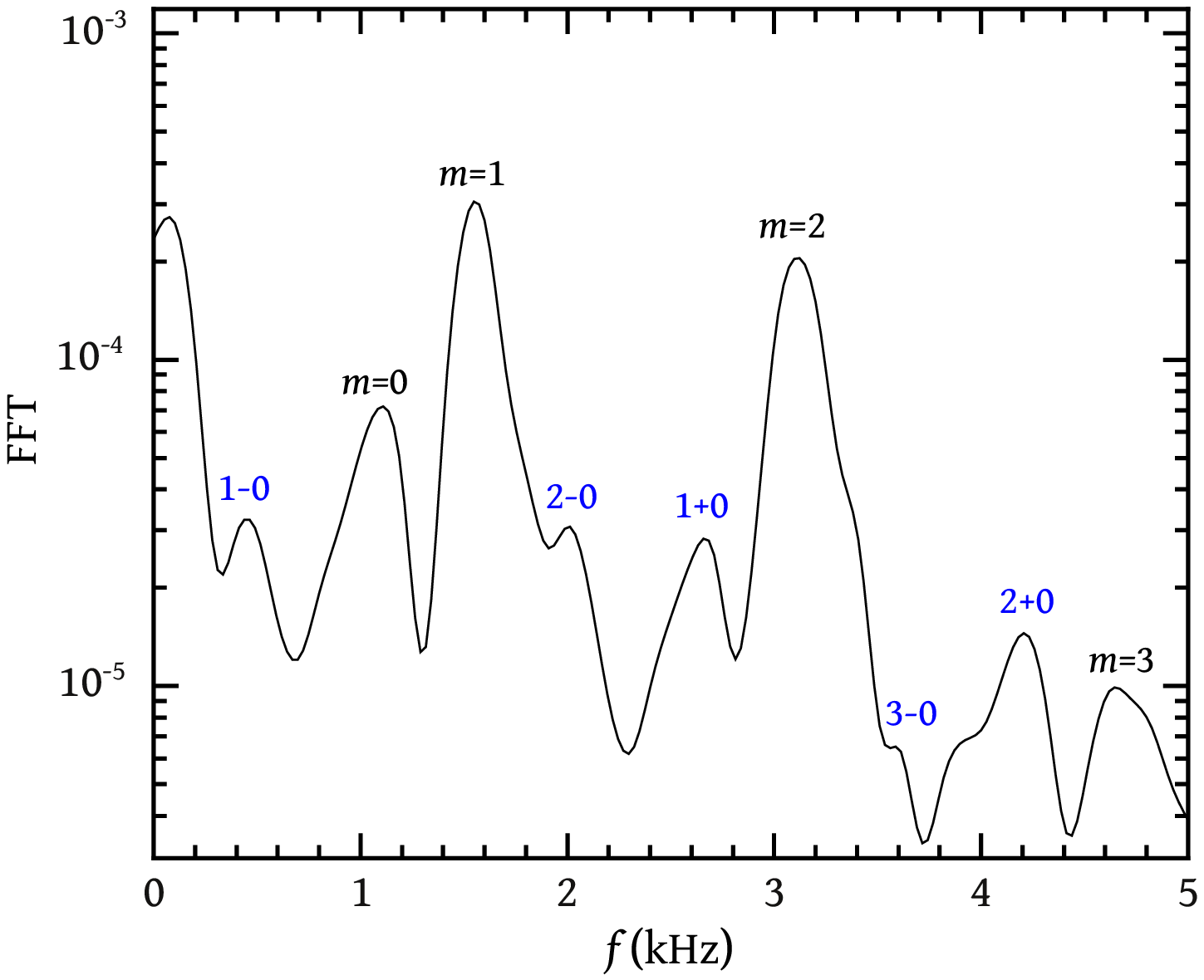}
  \caption{Same as Fig. \ref{fig:shen12135}, but for model LS 12135.}
  \label{fig:ls12135}
\end{figure}

The unequal-mass model LS 12135 shows similar qualitative properties as the corresponding
Shen 12135 model, except that all mode frequencies are considerably higher. In addition, a larger
number of nonlinear components can be clearly identified (bottom panel of Fig. 
\ref{fig:ls12135}). The $f_-$ vs. ``2-0'' coincidence remains. The corresponding equal-mass model
LS 135135 shows only the $m=0$ and $m=2$ modes dominating, with the addition of the ``2-0'' and
``2+0'' combination frequencies. As in the previous models, there is again 
an $f_-$ vs. ``2-0'' coincidence.

For the strange matter EOS MIT60, we examined the equal-mass case MIT60 1111, where each 
individual component of the binary system has a mass of only 1.1 $M_\odot$ (Fig. 
\ref{fig:mit60-1111}). The reason is
that for an individual mass of 1.35 $M_\odot$, an equal-mass binary system forms a black hole
soon after merging. Notice that the density profile of the MIT60 models differs drastically
from the case of the hadronic EOSs. It is very flat and terminates at a high value at the
surface. Especially for the low-mass model of 1.1 $M_\odot$, the density profile is roughly 
uniform. This causes the oscillation properties of the MIT60 models to differ considerably
from those of the hadronic models. The GW spectrum of the MIT60 1111 model
is still qualitatively similar to the previous hadronic models and one can still identify
a triplet of frequencies $f_-$, $f_2$ and $f_+$. However, in this case, the frequency of the
quasiradial mode is higher than for the hadronic models and, in fact, coincides with 
the frequency of the ``2-0'' nonlinear component. The latter differs, for this model only, 
from the $f_-$ peak in the GW spectrum. It is possible that the 
 $f_-$, $f_2$ and $f_+$ triplet is caused by the nonlinear interaction of the $f_2$ mode with a 
mode other than the quasiradial mode or that these are combination frequencies of higher 
order. Notice also that, for this model, the frequency of the $m=2$ mode is twice the
frequency of the quasiradial mode. Such a coincidence, which leads to a resonance between
the two modes and to enhanced GW emission has also been observed in simulations of 
phase-transition-induced instabilities in rotating compact stars 
\citep{Abdikamalov2009,Dimmelmeier2009}.

For the unequal mass MIT60 12135 model (Fig. 
\ref{fig:mit60-1111}), the frequency of the quasiradial mode is smaller than for the
MIT60 1111 model, while the frequency of the $m=2$ mode is higher. As a consequence, 
the frequency of the ``2-0'' nonlinear component  is not far from the $f_-$ peak in the
GW spectrum, although the agreement is not as good as in the case of
the hadronic EOSs. One should keep in mind, however, that the post-merger remnant is 
evolving in time, contracting its radius and reaching higher densities. This is causing
the different oscillation frequencies to change continually in time. One cannot, therefore,
expect a perfect match between the measured peaks in the Fourier transform of evolved
variables and the nonlinear features of the GW spectrum. 

We thus find that the GW spectrum of the post-merger phase 
is characterized by a triplet of frequencies that clearly coincides with the frequencies 
of the $m=2$ oscillation mode of the fluid and its interaction with the quasiradial mode
in all cases with mass in the range 1.2 $M_\odot$ to 1.35 $M_\odot$. Only in the low-mass
MIT60 1111 case the triplet has a different origin. 

\section{Comparison to previous work}
\label{sec:comp}

\cite{Allen1999} presented an initial study of binary neutron star mergers as 
a linear perturbation problem of an isolated star (a so-called close-limit approximation, 
in analogy to the corresponding approximation in binary black hole mergers). 
They showed that several fluid modes would be excited, but their study, being linear, 
did not take into account nonlinear combination frequencies, as we do here in our
fully nonlinear approach. 

In \cite{Zhuge1994} the frequency peak $f_-$ in the GW spectrum was
associated with the rotating barlike structure formed immediately after merging, 
while it was suggested that the main $f_2$ peak is due to a low-order $p$-mode.
\cite{2002PhRvD..65j3005O} proposed that the $f_2$ peak, being the quadrupole
frequency of the post-merger remnant, could be used to distinguish a soft EOS from
a stiff EOS. \cite{Shibata2002} also suggested that the post-merger 
GW spectrum is produced by non-axisymmetric oscillation modes of the merged
object and arrived at empirical relations (for polytropic EOSs) for the two
frequencies $f_-$ and $f_2$, without identifying their origin. Based on the dependence
of these two frequencies on the parameters of the polytropic EOS, they suggested that 
these two frequencies could be used to constrain the stiffness of the EOS. 

In \cite{2005PhRvD..71h4021S} it was suggested that quasiradial oscillations
modulate the post-merger signal and it was noticed that the difference in frequency
between the two main peaks (i.e. the peaks we call $f_-$ and $f_2$ here) 
is approximately equal to the frequency of the quasiradial oscillation. Furthermore, 
in \cite{Shibata2006} and in \cite{Kiuchi2009} two strong sideband peaks 
(forming a triplet with the
$f_2$ peak) were pointed out to exist in some cases and were associated with a modulation 
by large-amplitude quasiradial oscillations. The results of our mode analysis are in line with
the above observations and expectations. 

In contrast, \cite{Baiotti2008} described the repeated appearance of a 
double core structure in a particular high-mass merged object as a dynamical 
bar-mode instability which develops and is quickly suppressed again several times. The 
double core structure appeared at a regular interval of 2 ms. According to our
current results, this should be simply the period of the quasiradial oscillation and
instead of a dynamical instability the modulation is due to the ``2-0'' nonlinear
component (in other, low-mass models presented in \cite{Baiotti2008}, a 
bar persists for a long time and a possible association to a dynamical instability is
worth further investigation). 

\begin{figure} 
  \includegraphics[width=8.1cm]{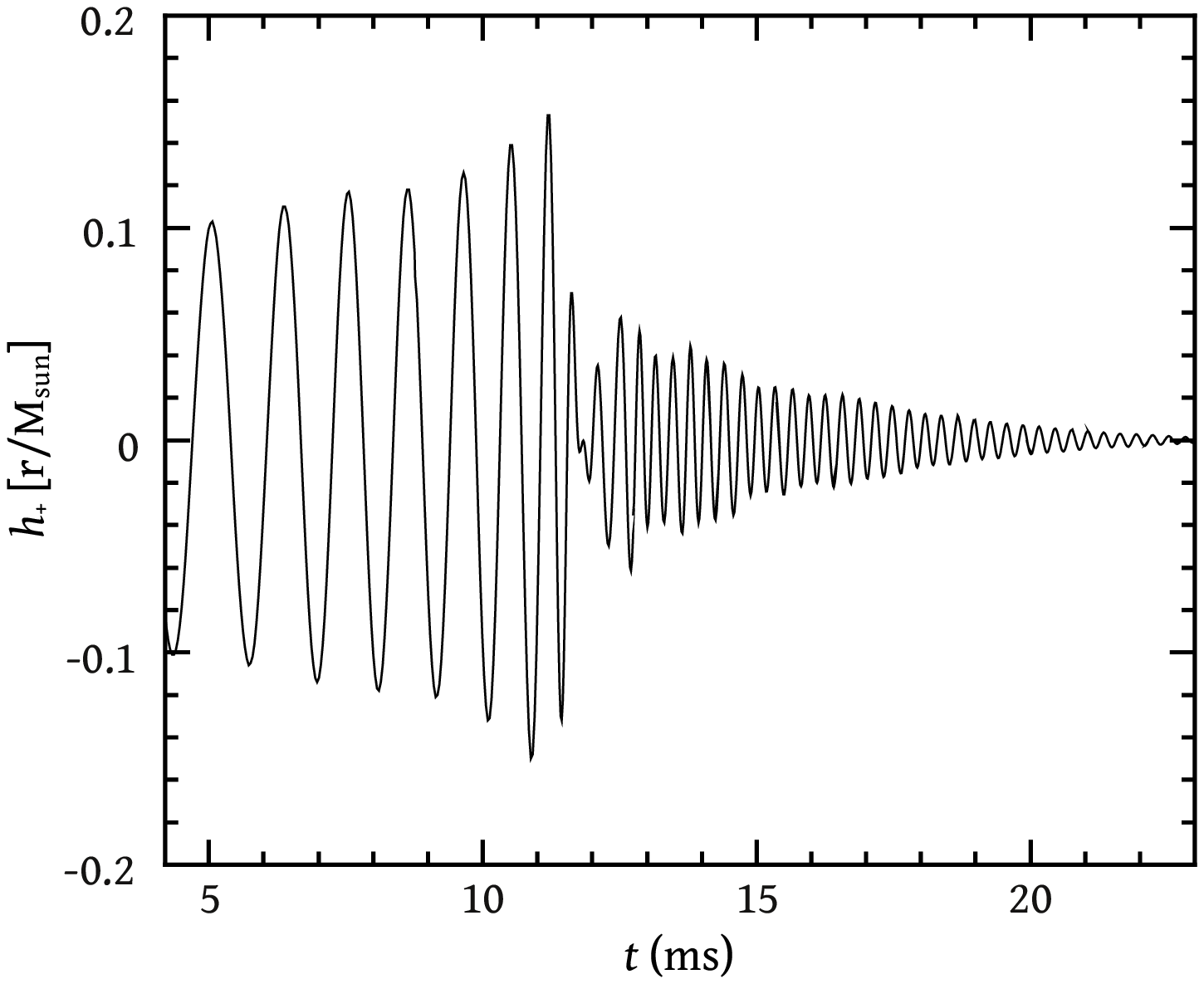}
\includegraphics[width=8.1cm]{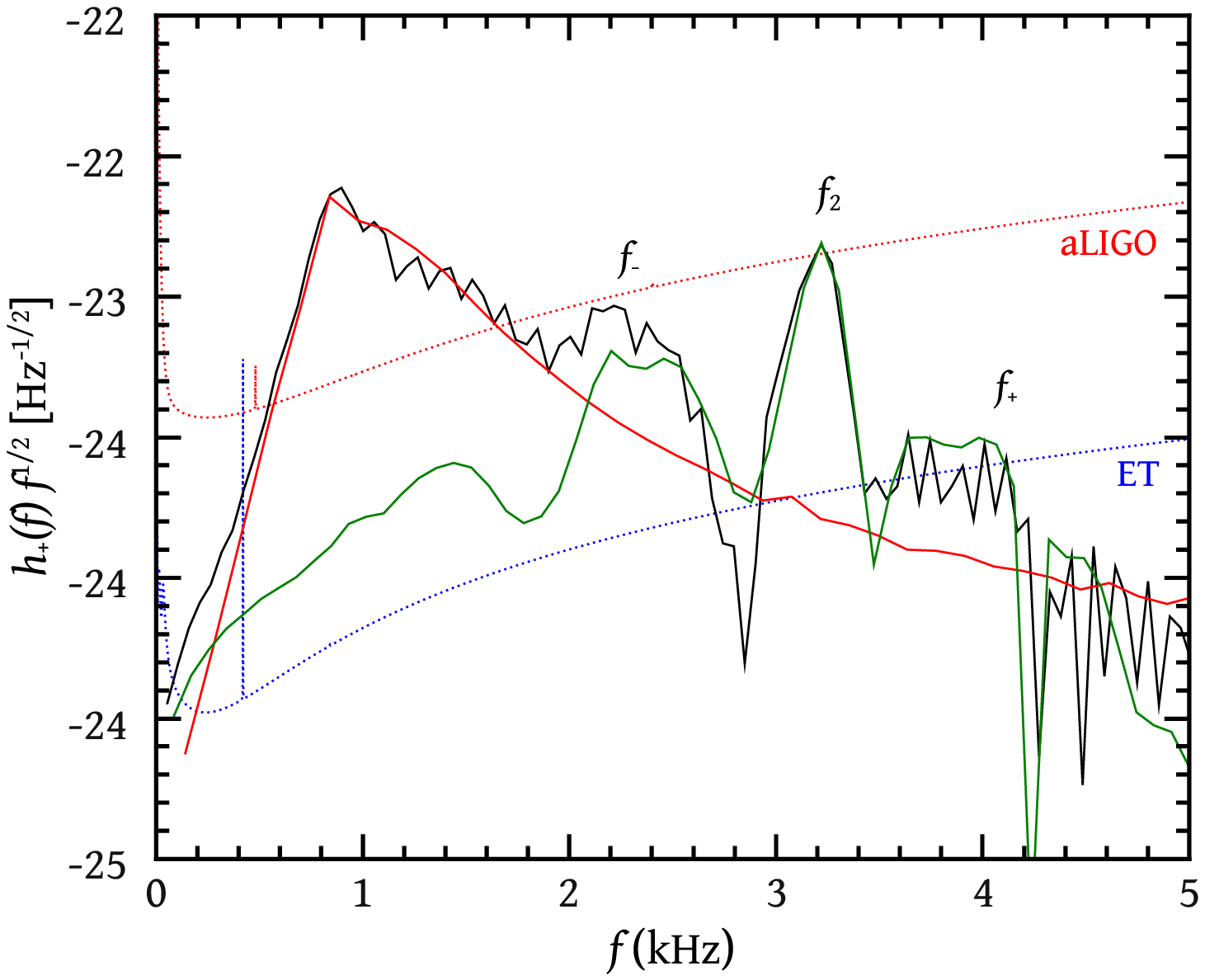}
\includegraphics[width=8.1cm]{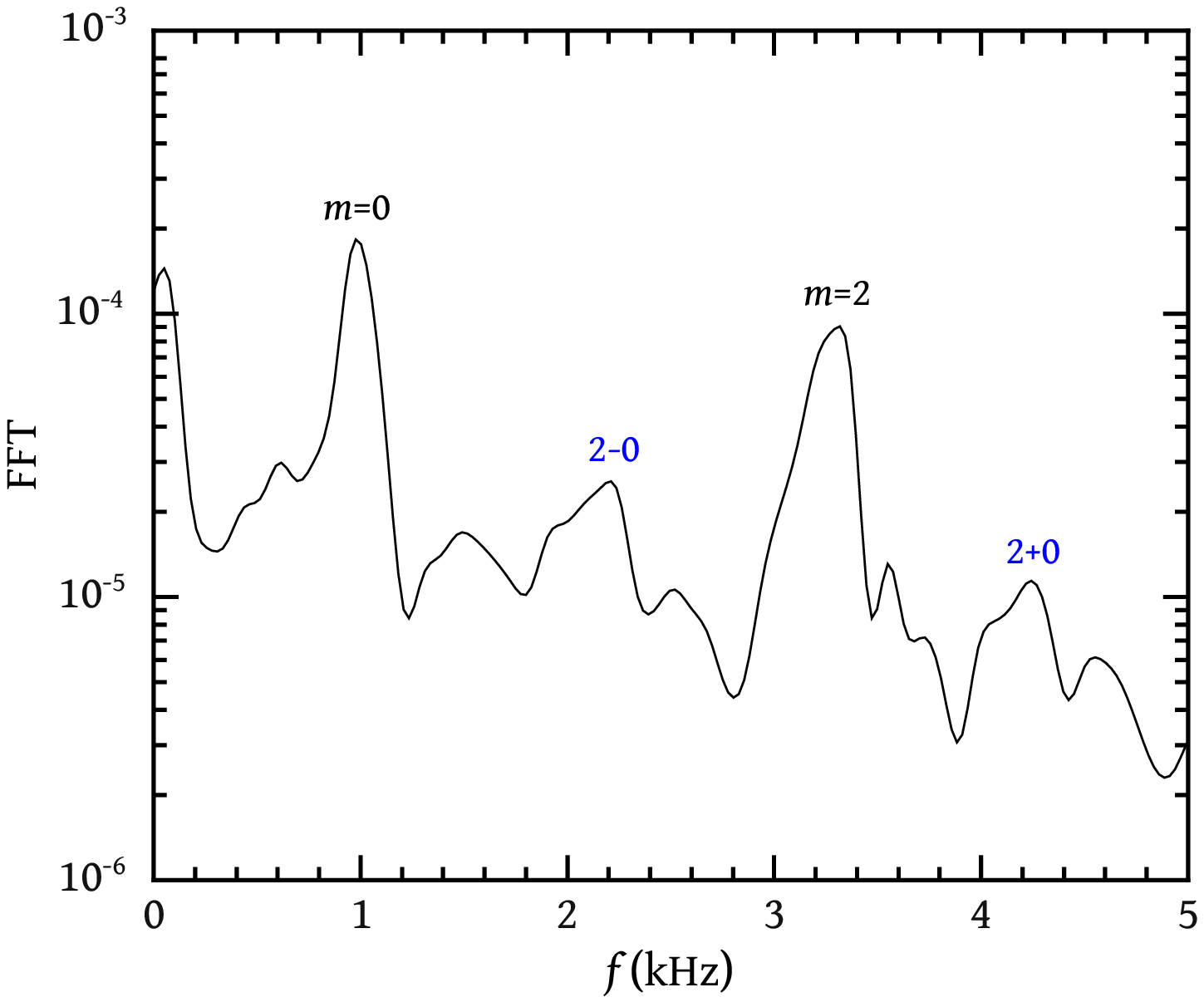}
  \caption{Same as Fig. \ref{fig:shen12135}, but for model LS 135135.}
  \label{fig:ls135135}
\end{figure}

\begin{figure} 
  \includegraphics[width=8.1cm]{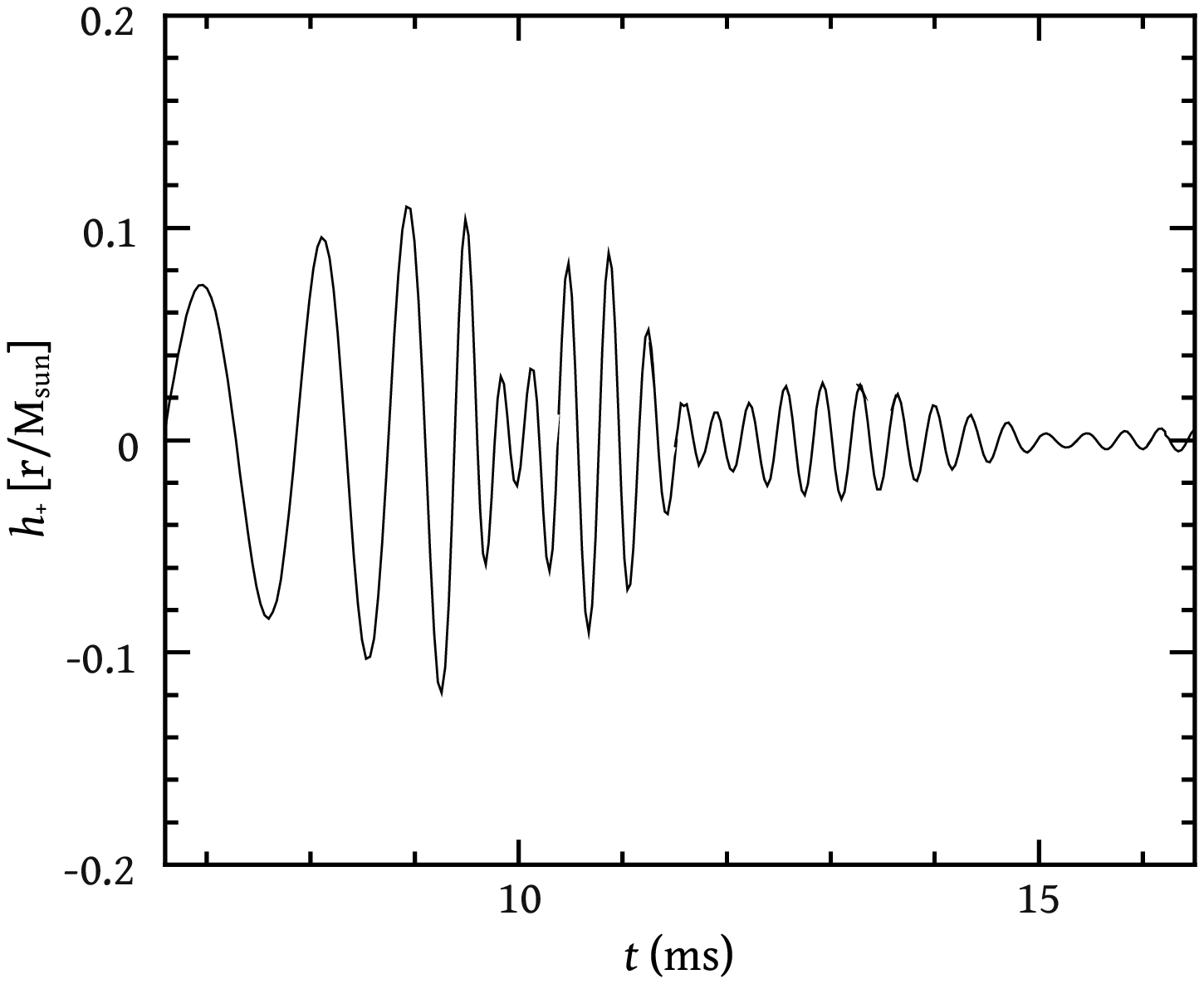}
 \includegraphics[width=8.1cm]{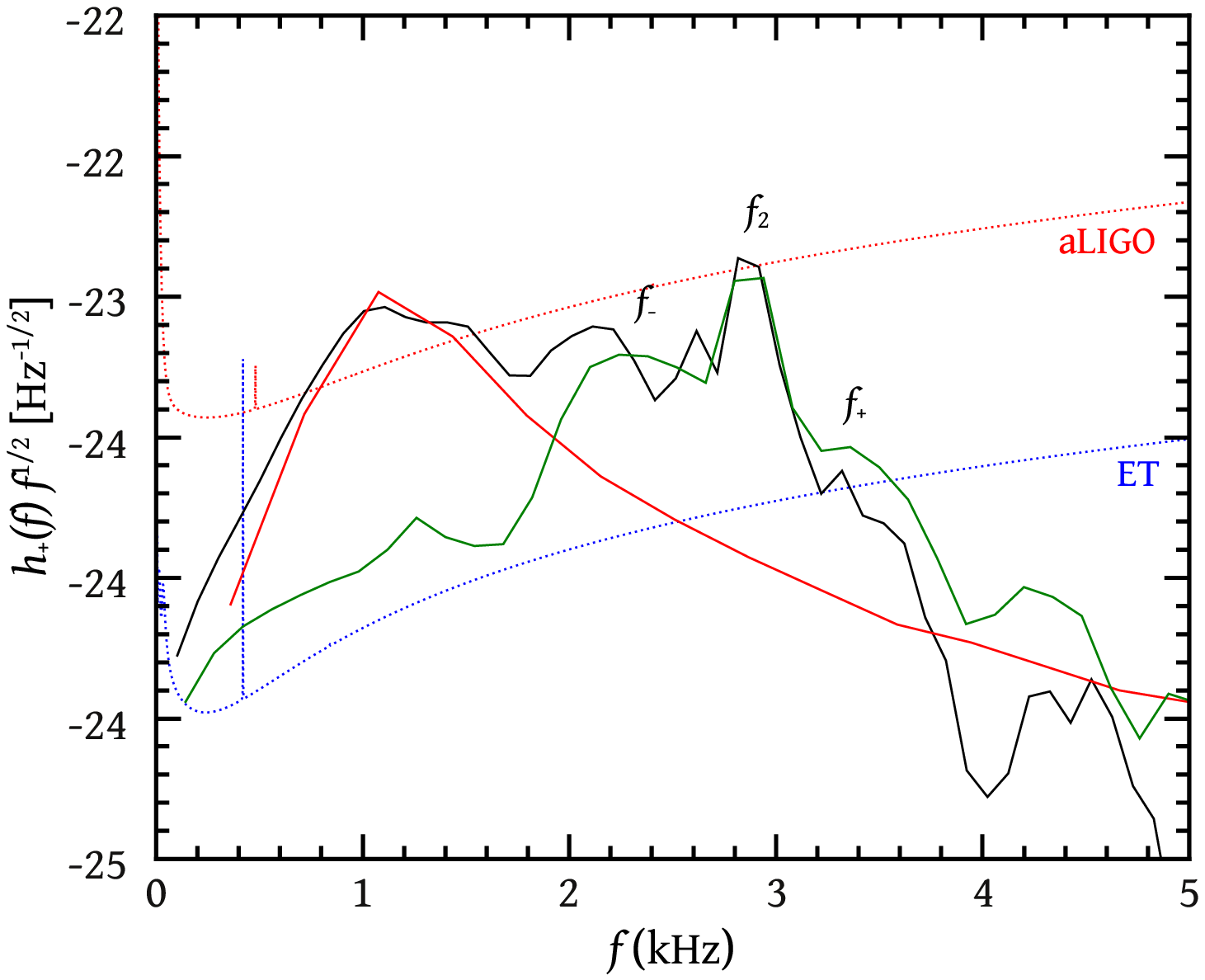}
\includegraphics[width=8.1cm]{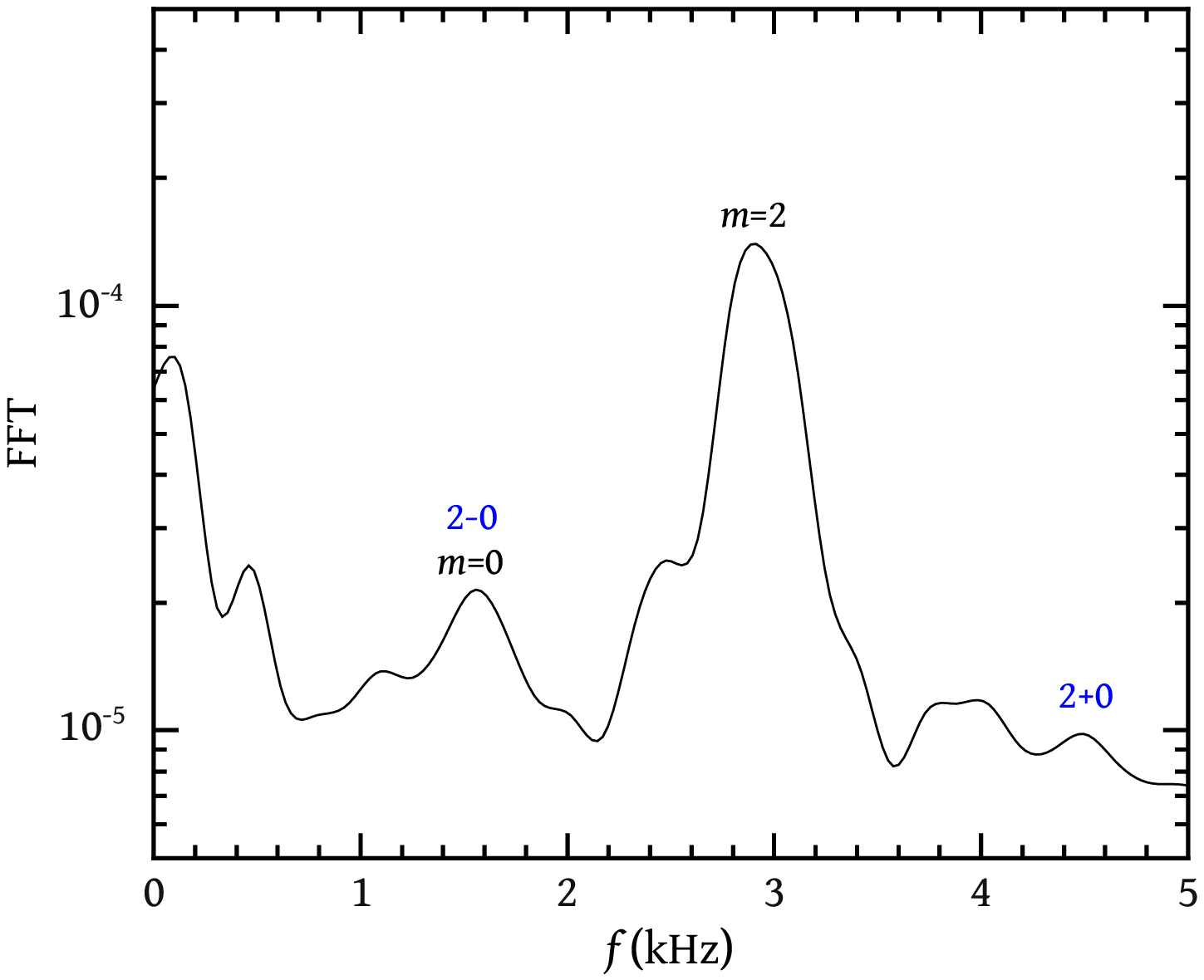}
  \caption{Same as Fig. \ref{fig:shen12135}, but for model MIT60 1111.}
  \label{fig:mit60-1111}
\end{figure}

\begin{figure}
  \includegraphics[width=8.1cm]{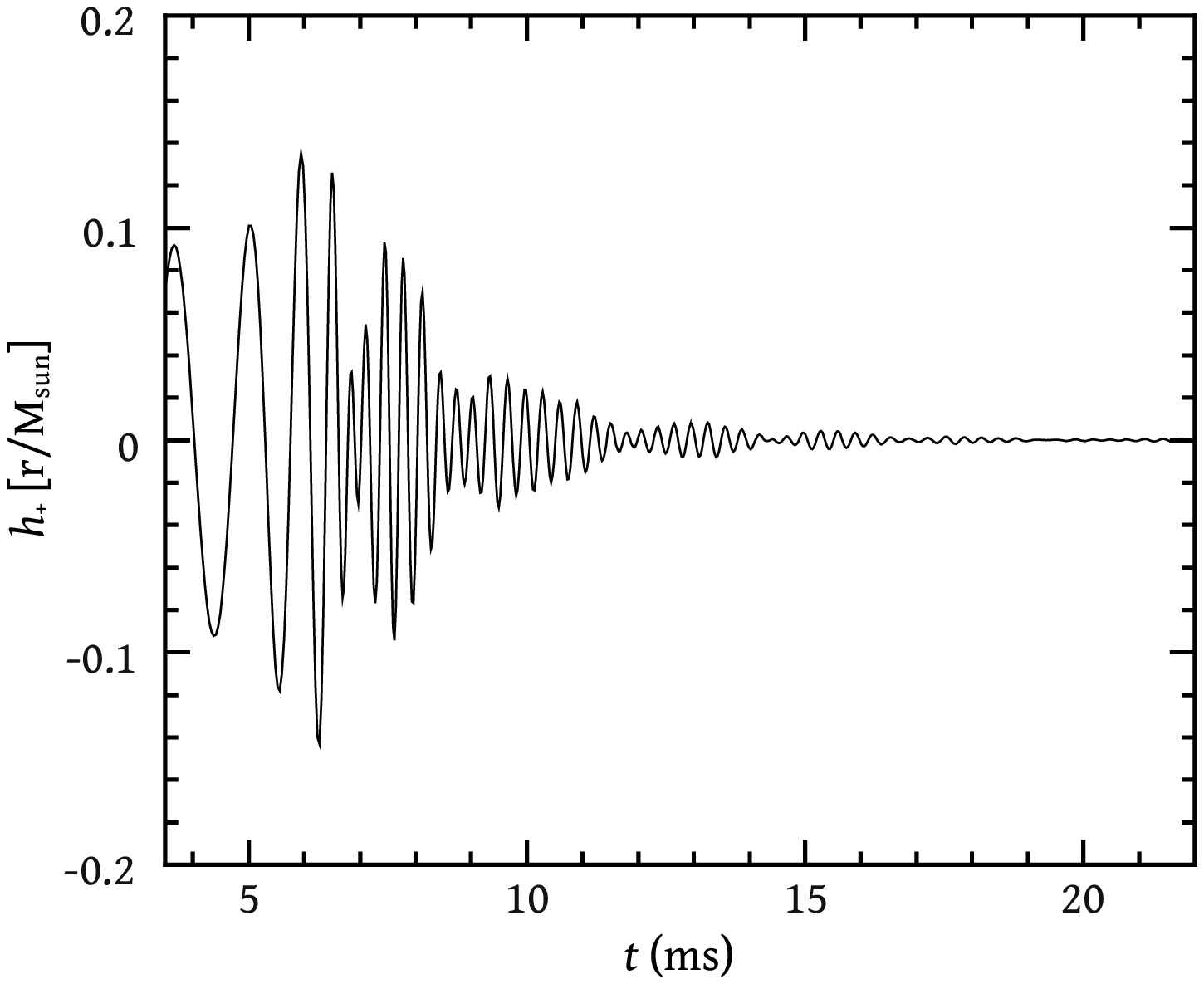}
 \includegraphics[width=8.1cm]{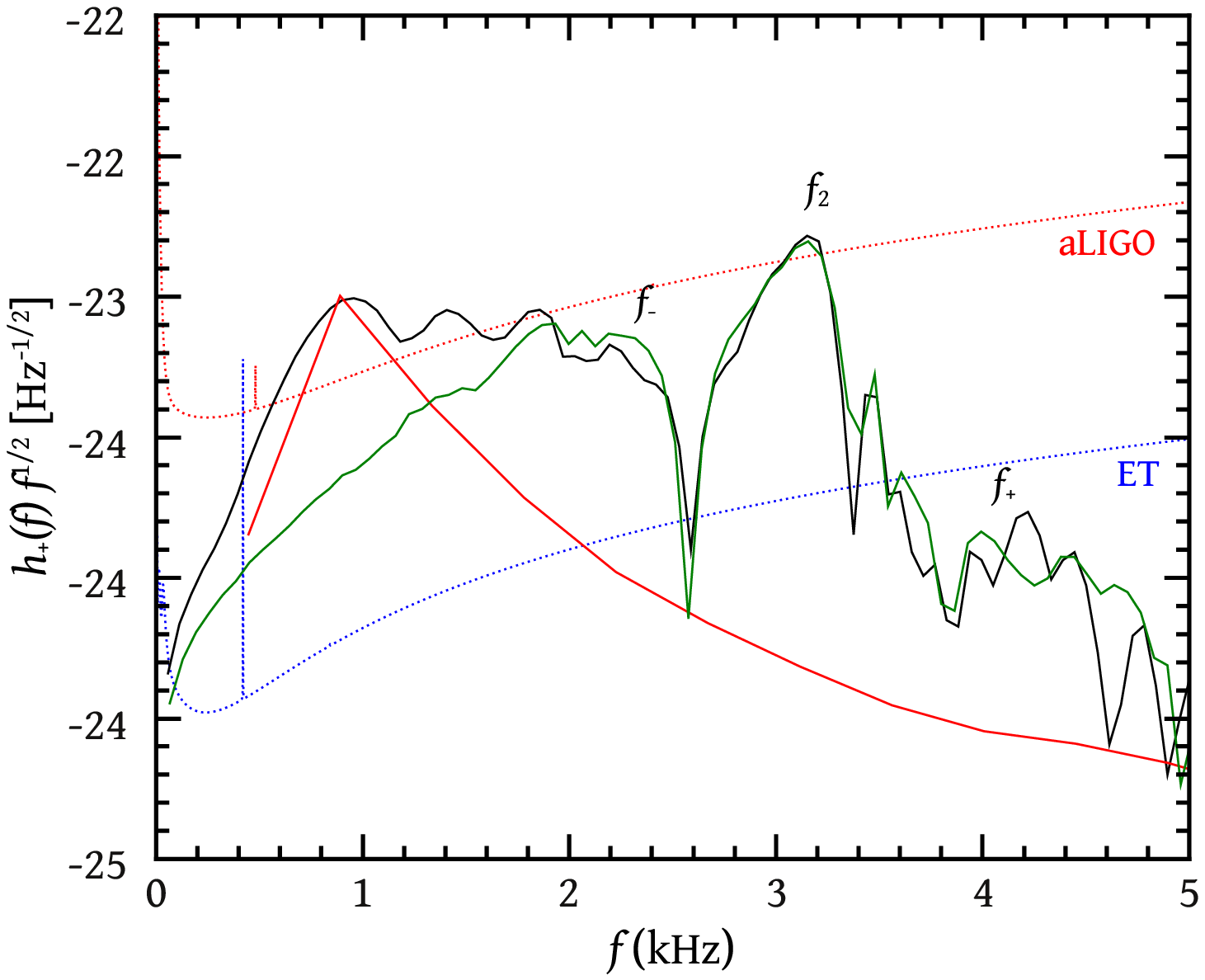}
 \includegraphics[width=8.1cm]{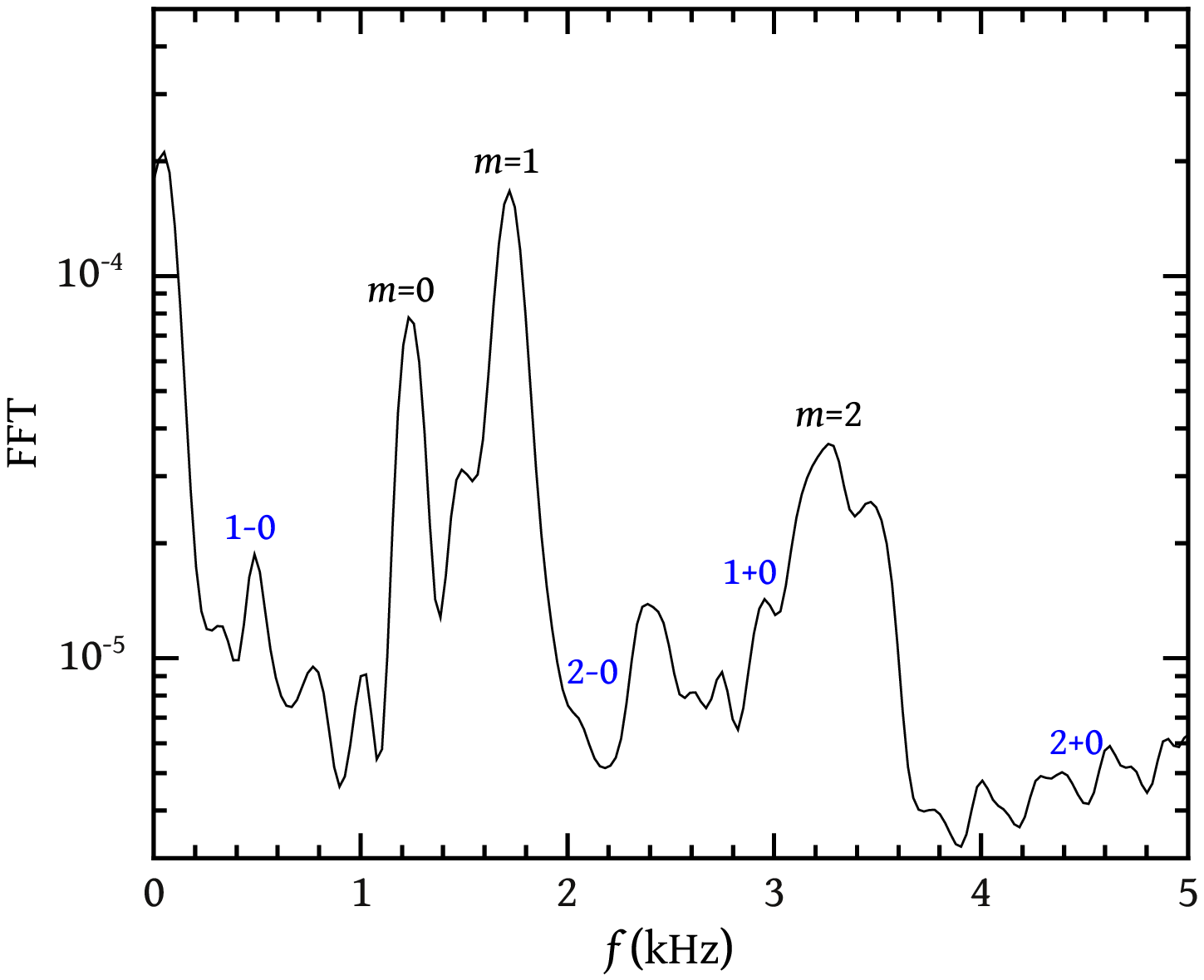}
  \caption{Same as Fig. \ref{fig:shen12135}, but for model MIT60 12135}
  \label{fig:mit60-12135}
\end{figure}

\section{Conclusions}
\label{sec:dis}

We have studied the excitation of nonaxisymmetric oscillations in the post-merger
phase of binary compact object mergers. Our analysis is based on general-relativistic 
simulations, using SPH for the evolution of matter, and we
used a set of equal-mass and unequal-mass models, described by two
nonzero-temperature hadronic EOSs and by one strange star 
EOS. We studied the oscillations through Fourier transforms of the evolved matter 
variables and identified a number of oscillation modes, as well as several  
nonlinear components (combination frequencies). The dominant 
$m=2$ mode forms a triplet with two nonlinear components that are the result of a coupling
to the quasiradial mode. A corresponding triplet of frequencies was identified in the 
GW spectrum, when the individual masses of the compact objects 
are in the most likely range of 1.2 to 1.35 $M_\odot$. A specific frequency
peak in the GW spectrum can thus be associated with the nonlinear 
component resulting from the difference between the $m=2$ mode and the quasiradial mode. 
This association is especially strong in the case of hadronic EOSs and could
be exploited in the case of future detections, in order to characterize the
properties of the post-merger remnant. For this, it will be necessary to 
obtain accurate frequencies of the quasi\-radial and quadrupole oscillation 
modes for a large sample of theoretically possible post-merger remnants and construct 
empirical relations, depending on a few gross properties, such as the mass and
radius of the star. Given the determination of two frequencies and of the total
mass of the system from the inspiral signal, these empirical relations could
be inverted to yield crucial information on the properties of high-density 
relativistic objects.   
 
It would further be interesting to test the sensitivity of our results to the 
spatial conformal flatness approximation employed here, as well as to assess the
influence of magnetic fields on the gravitational wave spectrum. The latter 
effect can be expected to be small for realistic magnetic fields, although there 
could be a region in the mass vs. magnetic field parameter space in which the 
effect could be measurable (for recent results, see \cite{Giacomazzo2011}).

\section*{Acknowledgements}
We are indebted to Harry Dimmelmeier and John L. Friedman for useful discussions 
and to Luciano Rezzolla for useful comments on the manuscript. 
We thank Stefan Hild for providing the sensitivity curve of the Einstein Telescope 
and David Shoemaker for providing the Advanced LIGO sensitivity curve and for reading 
the manuscript. In Garching the project was supported by the Deutsche Forschungsgemeinschaft
through the Transregional Collaborative Research Centers SFB/TR~7
``Gravitational Wave Astronomy'' and SFB/TR~27 ``Neutrinos and Beyond'',
and the Cluster of Excellence EXC~153 ``Origin and Structure of the Universe''
({\tt http://www.universe-cluster.de}).
Computer time grants at the Leib\-niz-Re\-chen\-zentrum
M\"unchen, and the RZG in Garching are acknowledged. 
This work was supported by CompStar, a Research Networking Programme of the 
European Science Foundation.

\bsp

\label{lastpage}

\bibliographystyle{mn2e} 
\bibliography{references}

\end{document}